\documentclass[11pt,a4paper]{article}

\usepackage[margin=1in]{geometry}
\usepackage{graphicx}
\usepackage{xcolor}
\usepackage{tocloft}
\usepackage{enumitem}
\usepackage{setspace}
\usepackage{titlesec}
\usepackage{amssymb}
\usepackage[unicode=true]{hyperref}
\usepackage[super, square, sort&compress, numbers]{natbib}
\usepackage{amsmath}
\usepackage{graphicx}
\usepackage{tabularx}
\usepackage{booktabs}
\usepackage{epsfig}
\usepackage{ulem}

\hypersetup{
    colorlinks=true,
    linkcolor=blue,
    bookmarksopen=true,
    bookmarksnumbered=true
}

\titleformat{\section}{\large\bfseries}{}{0em}{}
\titleformat{\subsection}{\bfseries}{}{0em}{}

\begin{document}

\begin{center}
{\Large\bfseries Interlayer pairing mechanism for bilayer nickelate superconductors}
\end{center}
\vspace{0.6em}
\begin{center}
{\normalsize Jiangfan Wang,\textsuperscript{1,*} Yi-feng Yang,\textsuperscript{2,3,*}}
\end{center}
\vspace{0.4em}
\begin{center}
{\small
\textsuperscript{1}School of Physics, Hangzhou Normal University, Hangzhou, Zhejiang 311121, China\\
\textsuperscript{2}Beijing National Laboratory for Condensed Matter Physics and Institute of Physics,
Chinese Academy of Sciences, Beijing 100190, China\\
\textsuperscript{3}School of Physical Sciences, University of Chinese Academy of Sciences, Beijing 100049, China\\
\textsuperscript{*}Correspondence: jfwang@hznu.edu.cn (J.W.); yifeng@iphy.ac.cn (Y.-f.Y.)
}
\end{center}

\vspace{0.8em}

\begin{abstract}
\noindent The discovery of superconductivity in Ruddlesden-Popper bilayer nickelates under both high pressure and ambient conditions has opened a new paradigm for exploring unconventional superconductivity. This review provides a brief survey of theoretical progress on bilayer nickelate superconductors. Drawing from the key experimental observations, we summarize essential physical ingredients including the hybridized Ni-3$d_{x^2-y^2}$ and 3$d_{z^2}$ electronic structure, orbital-dependent electronic correlation, Hund's coupling, and strong interlayer magnetic coupling. The fundamental theoretical models including the bilayer two-orbital Hubbard model and its minimal $t$-$J$ variants are introduced. Starting from the atomic-limit interlayer valence bond picture of the half-filled $d_{z^2}$ orbital, we elaborate on strong correlation interlayer pairing mechanisms based on different limiting considerations. Specific emphasis is placed on the hybridization mechanism, where the $d_{z^2}$ local singlet pairs provide the pairing energy and their hybridization with itinerant $d_{x^2-y^2}$ promotes superconducting phase coherence. We further analyze the pairing symmetry, the dependence of $T_c$ on various internal and external parameters, the nontrivial normal state properties including the Fermi liquid, non-Fermi liquid, weakly insulating and pseudogap behaviors. Effects of pressure tuning, oxygen content, and Kondo scattering induced by oxygen vacancies are also discussed. Finally, weak correlation theories based on spin fluctuations associated with Fermi surface nesting are briefly covered.
\end{abstract}
\begin{center}
\end{center}
\vspace{1.5em}

\section{I. INTRODUCTION}

The pursuit of unconventional superconductivity has long been guided by the paradigm of the layered copper oxides, \cite{Keimer2015,Scalapino2012cuprate,Wang2023BSCCO} where the Cu$^{2+}$ $3d^9$ electronic configuration gives rise to a spin-1/2 Mott insulator at half filling and induces superconductivity upon doping. The quest for analogous systems beyond copper oxides led to the discovery of the infinite-layer nickelate superconductor Nd$_{0.8}$Sr$_{0.2}$NiO$_2$, \cite{Li2019,JGCheng2022Pressure,Hwang2023Linear,Fan2024Capping,XJZhou2024Magnetic,Ariando2025Bulk,QSWang2026Persistent,DFLi2026,Hwang2026Nd112,LSun2026Enhanced} where Ni$^{1+}$ adopts a similar $3d^9$ configuration. The compound belongs to the reduced Ruddlesden-Popper (RP) perovskite family, which is obtained from the RP phase $R_{n+1}$Ni$_n$O$_{3n+1}$ ($R$: rare-earth elements) by removing the apical oxygen ions from the NiO$_6$ octahedra. This breakthrough ignited efforts to explore other RP nickelates for potential high-temperature superconductivity.

A pivotal moment arrived in 2023 with the discovery of high-temperature superconductivity in the bilayer RP nickelate La$_{3}$Ni$_{2}$O$_{7}$ under high pressure. \cite{MWang2023Nature,JGCheng2023,HQYuan2024,ZChen2024,JGCheng2024Nature,HHWen2024,HHWen2025Evolution,XJZhou2024,YWang2025Interstitial,LShu2024,MWang2023SDW,XHChen2025SDW,Guguchia2025SDW,DLFeng2024,MWang2024,HLuo2026spin,MWang2024b,LYang2024,Shen2025,JJZhang2025,MWang2025,XHChen2025review,KeWang2026,GHCao2026,JZhao2026,Talantsev2024,Li2025SciBull,DLFeng2026NC,JZhao2026Interlayer,MWang2026Sm,MWang2026Regulating,HHWen2025Andreev,XLHuang2025Revealing,HkMao2025Direct,JingGuo2026} The transition temperature $T_c$ reaches about 80 K at 14 GPa, making bilayer nickelates the second correlated electron system after cuprates to have a $T_c$ above the boiling point of liquid nitrogen. Unlike cuprates, the bilayer nickelate has a nominal Ni$^{2.5+}$ valence, corresponding to an averaged  $3d^{7.5}$ configuration with active $d_{z^2}$ and $d_{x^2-y^2}$ orbitals around the Fermi energy. \cite{MWang2023Nature} The two NiO$_6$ octahedra share the same inner apical oxygen along the $c$-axis, leading to strong interlayer coupling between the nearly half-filled Ni-3$d_{z^2}$ orbital. This strong interlayer coupling was confirmed by resonant inelastic X-ray scattering (RIXS) \cite{DLFeng2024} and inelastic neutron scattering (INS) measurements, \cite{MWang2024,HLuo2026spin,JZhao2026} which is essential for both superconductivity and the density waves observed at ambient pressure. \cite{MWang2023Nature,LShu2024,MWang2023SDW,XHChen2025SDW,Guguchia2025SDW,DLFeng2024,MWang2024} Another important feature is the strange metallic linear-in-$T$ behavior of the resistivity above optimal $T_c$, \cite{HQYuan2024,Shen2025} which suggests that strong electronic correlation plays an important role in this system.  Soon after the discovery of the bilayer nickelate superconductor, the trilayer nickelate La$_4$Ni$_3$O$_{10}$ was also found to be superconducting under pressure, but with a much lower transition temperature $T_c\sim$ 30 K. \cite{Sakakibara2024,HHWen2024trilayer,JZhao2024,YPQi2025PRX,JZhao2025PRX,Kim2025Orbital,NLWang2025Collapse,Guguchia2026Pressure,LXYang2025Dichotomy,MWang2026Distinct} The reduced $T_c$ in trilayer nickelate contributes another important difference from the cuprates, \cite{QQin2024,YFYang2024} of which the trilayer compounds typically have the highest $T_c$ among all layered cuprates. \cite{Scalapino2012cuprate,Wang2023BSCCO} 

The high-pressure condition required for nickelate superconductivity strongly hindered experimental progress in these compounds. In early 2025, the field was significantly advanced by the discovery of ambient-pressure superconductivity with onset $T_c\sim 40$ K in compressively strained La$_3$Ni$_2$O$_7$ or (La,$R$)$_3$Ni$_2$O$_7$ ($R=$ Pr, Nd, Sm, Sr) thin films grown on LaSrAlO$_4$ (LSAO) substrates, \cite{Hwang2025,Chen2025,Hwang2025b,QKXue2025,ZXShen2025,Hwang2025FL,HHWen2025,ZYChen2025,ZYChen2025SIT,Nie2025Sr,Nie2025PRL_crossover,HHWen2026pressure,HHWen2026Heavily,ZYChen60K,ZYChen2026,Hwang2025Reducing,Nie2025ARPES,HHWen2024Kondo,Kumar2026,ZYChen2026Three,Hwang2026halfdome,KuiJin2026,KJZhou2026,Goodge2026Structural,XLu2026Doping,QKXue2026STM,WLee2026Interlayer,Osada2026} eliminating the need for external pressure and greatly expanding the experimental accessibility of these materials. Later,  La$_2$PrNi$_2$O$_7$ thin films grown on LaAlO$_3$ (LAO) substrates with less compressive strain were reported to be superconducting at ambient pressure, but with a much lower $T_c\sim 10$ K. \cite{Hwang2025Reducing} In addition, La$_3$Ni$_2$O$_7$ thin films grown on NdGaO$_3$ and SrTiO$_3$ substrates \cite{Tsukazaki2025Strain} and Pr$_3$Ni$_2$O$_7$ thin films on LAO substrate \cite{Tsukazaki2026Pressure} were also found to be superconducting under high hydrostatic pressures. More recently, La$_2$PrNi$_2$O$_7$ films on NdAlO$_3$ substrates were reported to exhibit a high $T_c\sim 60$ K with an extreme compressive strain (-2.14\%) \cite{XHChen2026Electron,XHChen2026Thermo}. As material growth techniques continuously improve, $T_c$ now reaches nearly 100 K in pressurized bulk \cite{JJZhang2025,MWang2025} and over 60 K in compressively strained thin films. \cite{ZYChen60K,Tsukazaki2026Pressure,XHChen2026Electron,XHChen2026Thermo,Osada2026} This rapid evolution has established the RP nickelates as a compelling new platform for studying high-temperature superconductivity.

The rapid experimental progress in bilayer nickelate superconductors has been paralleled by extensive theoretical investigations. Based on the angle-resolved-photoemission spectroscopy (ARPES) measurements and first-principles electronic structure calculations, \cite{Dagotto2023,Dagotto2024,Werner2023,Leonov2023,Hirschfeld2024,YYCao2024,Eremin2024,BHuang2024,ZYLu2024EPC,Botana2025electronic,CJia2025,DXYao2025CPL,DXYao2025film,DXYao2025Theorectical,Verraes2025,GSu2025Unifying,ZZhang2026PRL,DfDuan2026} it is generally accepted that the low-energy description of bilayer nickelates should involve both the Ni 3$d_{x^2-y^2}$ and 3$d_{z^2}$ orbitals on the bilayer structure, which motivates a large number of theoretical studies based on different bilayer two-orbital models. \cite{YFYang2023,QQin2023,GMZhang2023,DXYao2023,FYang2023,QHWang2023,WQChen2025,JPHu2025,QHWang2025,JPHu2025PRL,JPHu2025Opposite,Dagotto2025RPA,QHWang2026Strain,QHWang2026Tunable,Kuroki2024,Eremin2023,HHChen2025,Savrasov2024,Braz2025PRR,FYang2025Band,Watanabe2026VMC,Kuroki2024Pair,DXYao2025DMRG,QHWang2025VMC,Dagotto2026Interlayer,WWu2025,WWu2025Intertwined,ZYLu2024,ZYLu2024Correlation,ZYLu2025DMFT,JXLi2025,HHChen2026Nearly,QQin2025,Wang2025,Wang2026,Wang2026unified,YFYang2025CPL,CJWu2024CPL,CJWu2024,CJWu2024PRL,GSu2024,YHZhang2023,YHZhang2025,GSu2025,CJWu2025,FYang2025Pairing,FYang2026Unified,ZYWeng2024,FWang2024CPL,DXYao2023tJ,DXYao2025pairing,YZYou2023,WKu2024,Khaliullin2026,HLiu2026triplon,DXYao2025Nd,WLi2024,Kuroki2025tJ,Bohrdt2024DMRG,Bohrt2024Feshbach,Kuroki2024DMRG,FYang2026NC,MJiang2026EleDope,TZhou2023Impurity,Raghu2026} Among these, the bilayer two-orbital Hubbard model has been intensively studied, \cite{DXYao2023,FYang2023,QHWang2023,WQChen2025,JPHu2025,QHWang2025,JPHu2025PRL,JPHu2025Opposite,Dagotto2025RPA,QHWang2026Strain,QHWang2026Tunable,Kuroki2024,Eremin2023,HHChen2025,Savrasov2024,Braz2025PRR,FYang2025Band,Watanabe2026VMC,Kuroki2024Pair,DXYao2025DMRG,QHWang2025VMC,Dagotto2026Interlayer,WWu2025,WWu2025Intertwined,ZYLu2024,ZYLu2024Correlation,ZYLu2025DMFT,JXLi2025,HHChen2026Nearly} which combines the tight-binding Hamiltonian from first-principles calculations with the multi-orbital Coulomb and Hund's interactions. In strong correlation regime, the nearly half filled $d_{z^2}$ electrons form local moments, which exhibit strong interlayer superexchange via the shared inner apical oxygen, leading to different $t$-$J$ like models. \cite{YFYang2023,QQin2023,GMZhang2023,QQin2025,Wang2025,Wang2026,Wang2026unified,YFYang2025CPL,CJWu2024CPL,CJWu2024,CJWu2024PRL,GSu2024,YHZhang2023,YHZhang2025,GSu2025,CJWu2025,FYang2025Pairing,FYang2026Unified,ZYWeng2024,FWang2024CPL,DXYao2023tJ,DXYao2025pairing,YZYou2023,WKu2024,Khaliullin2026,HLiu2026triplon,DXYao2025Nd,WLi2024,Kuroki2025tJ,Bohrdt2024DMRG,Bohrt2024Feshbach,Kuroki2024DMRG,FYang2026NC} Various theoretical approaches have been used to solve these models, which can roughly be classified into two classes --- weak correlation theories and strong correlation theories. The weak correlation theories treat both $d_{z^2}$ and $d_{x^2-y^2}$ as itinerant orbitals with well defined Fermi surfaces. By including electronic correlations in some perturbative ways such as the random-phase approximation (RPA) \cite{DXYao2023,FYang2023,Dagotto2024,Dagotto2025RPA,HHChen2025,Savrasov2024,Eremin2023,Braz2025PRR,FYang2025Band,WQChen2025,Raghu2026,Ryee2025} or functional renormalization group (FRG), \cite{QHWang2023,QHWang2025,QHWang2026Strain,QHWang2026Tunable,JPHu2025,JPHu2025PRL,JPHu2025Opposite} these methods then derive the strongest spin fluctuations associated with certain Fermi surface nestings, which provide the pairing glue and induce superconductivity. On the other hand, the strong correlation theories start with nearly localized $d_{z^2}$ spins with strong interlayer superexchange, and by considering their coupling with the more itinerant $d_{x^2-y^2}$ electrons  through the interorbital hybridization \cite{YFYang2023,QQin2023,GMZhang2023,QQin2025,Wang2025,Wang2026,Wang2026unified,YFYang2025CPL,WWu2025,Watanabe2026VMC} or Hund's coupling, \cite{CJWu2024CPL,CJWu2024,CJWu2024PRL,GSu2024,YHZhang2023,YHZhang2025,GSu2025,CJWu2025,FYang2025Pairing,FYang2026Unified,ZYWeng2024} the system forms interlayer Cooper pairs and achieves superconductivity.

This paper aims to provide a brief review of the interlayer pairing mechanisms of bilayer nickelates, focusing on the strong correlation theories. We start by summarizing the key ingredients from important experimental results in Section II, which allows us to distill the minimal effective model. In Section III, we elaborate on strong correlation interlayer pairing theories by considering different limiting cases of the effective model. Specific emphasis is placed on the two-component theory, in which the $d_{z^2}$ local singlet pairs provide the pairing energy and their hybridization with itinerant $d_{x^2-y^2}$ promotes superconducting phase coherence. We then analyze the pairing symmetry, the variation of $T_c$ with various parameters, the nontrivial normal state properties including the Fermi liquid (FL), non-Fermi liquid (NFL), weakly insulating and pseudogap behaviors. After that, we briefly discuss the effects of pressure tuning, oxygen content, and Kondo scattering induced by oxygen vacancies. Due to the limited length, weak correlation theories will be briefly covered at the end of Section III. Finally, Section IV gives our summary and outlook.

\section{II. KEY EXPERIMENTS AND EFFECTIVE MODELS}

We first summarize the most important experimental observations for both bulk and thin film bilayer nickelates, which enable us to extract key physical ingredients of these systems and obtain minimal effective models.  

\begin{itemize}[leftmargin=2.0em]
	\item \textit{Electronic band structure.---}Both the bulk and thin film bilayer nickelates have multi-orbital electronic structures consisting of the nickel $3d_{z^2}$ and $3d_{x^2-y^2}$ bands around the Fermi energy. Density-functional theory (DFT) calculations reveal strong interorbital hybridization and large vertical hopping of the $d_{z^2}$ orbital. \cite{MWang2023Nature,DXYao2023} The latter causes a large bonding-antibonding splitting of the hybridized bands, leading to a bonding $\alpha$ and an antibonding $\beta$ Fermi sheet with mixed $d_{x^2-y^2}$ and $d_{z^2}$ characters, and a $d_{z^2}$ dominated $\gamma$ bonding band near the Fermi energy. \cite{DXYao2023} For bulk La$_3$Ni$_2$O$_7$, both ARPES and DFT calculations show that the $\gamma$ band lies below the Fermi energy at ambient pressure. \cite{MWang2023Nature,XJZhou2024} As pressure increases, a first-order structural phase transition from the orthorhombic $Amam$ phase to the nearly tetragonal $Fmmm$ phase takes place at around 14 GPa (note that another transition to the tetragonal $I4/mmm$ phase occurs at higher pressure \cite{MWang2024b}); the vertical Ni-O-Ni bonds become straight and the $\gamma$ hole pockets emerge around the Brillouin zone corner according to first-principles calculations. \cite{MWang2023Nature} For thin films, the Ni-O-Ni vertical bonds are straightened by the compressive strain at ambient pressure, \cite{GSu2025Unifying,Goodge2026Structural} resulting in the $I4/mmm$ space group with a similar electronic structure to that of the pressurized bulk. DFT calculations suggest that the compressive strain lowers the $d_{z^2}$ orbital energy levels, \cite{Botana2025electronic,CJia2025,GSu2025Unifying,ZZhang2026PRL} but the interfacial Sr diffusion from the LSAO substrate may lead to intrinsic hole doping and reconstruct the $\gamma$ hole pockets. \cite{Chen2025,DXYao2025CPL,DXYao2025film,DXYao2025Theorectical,GSu2025Unifying,WQChen2025} Experimentally, both the presence and absence of $\gamma$ pockets have been observed in ARPES measurements, \cite{ZXShen2025,QKXue2025} leading to the debate on whether the $\gamma$ hole pocket is necessary for the superconductivity. 
\end{itemize}

\begin{itemize}[leftmargin=2.0em]
	\item\textit{Electron interactions.---}The Coulomb repulsion and Hund's rule coupling strongly renormalize the quasiparticle band structure. Early DFT+DMFT calculations predict a mass renormalization factor of $\sim5$ for the $d_{z^2}$ orbital and $\sim3$ for the $d_{x^2-y^2}$ orbital. \cite{YYCao2024} This orbital-selective mass renormalization was confirmed by ARPES measurements at ambient pressure \cite{XJZhou2024} and other DFT+DMFT calculations. \cite{ZYLu2024,ZYLu2024Correlation} Strong electronic correlation has also been detected from optical spectroscopy measurements, suggesting that the system may be in the proximity of a Mott phase. \cite{HHWen2024,HHWen2025Evolution}  These observations suggest that the nearly half-filled $d_{z^2}$ orbital is strongly correlated and close to forming local moments, while the nearly quarter-filled $d_{x^2-y^2}$ is relatively weakly correlated and more itinerant. Due to the large vertical hopping, the $d_{z^2}$ spins must have a strong interlayer superexchange, which has been confirmed by magnetic excitation measurements. \cite{DLFeng2024,MWang2024,HLuo2026spin,JZhao2026,XLu2026Doping} In addition, transport measurements of non-superconducting La$_{3}$Ni$_2$O$_{7-\delta}$ thin films \cite{KuiJin2026,HHWen2024Kondo,Kumar2026} and polycrystalline bulk samples \cite{KeWang2026} observed logarithmic upturn of electric resistivity \cite{KuiJin2026,HHWen2024Kondo,Kumar2026,KeWang2026} and negative magnetoresistivity,  \cite{KuiJin2026,HHWen2024Kondo} indicating the existence of Kondo scattering effect. This further supports the strong hybridization between the itinerant $d_{x^2-y^2}$ orbital and localized $d_{z^2}$ electrons. 
\end{itemize}

\begin{itemize}[leftmargin=2.0em]
	\item\textit{Magnetic coupling.---}The magnetic excitations in bulk La$_3$Ni$_2$O$_7$ have been measured by RIXS at ambient pressure. \cite{DLFeng2024} By fitting the experimental data using a simple Heisenberg model, Ref. \cite{DLFeng2024} obtained an interlayer magnetic coupling $J_\perp S$ of about 60--70 meV, much larger than the intralayer couplings. Similar conclusions have been obtained from INS measurements on  La$_3$Ni$_2$O$_{7-\delta}$ \cite{MWang2024,JZhao2026} as well as Pr and Nd doped samples. \cite{HLuo2026spin} For compressively strained La$_{3-x}$Sr$_x$Ni$_2$O$_7$ thin films, a recent RIXS measurement found magnetic modes with a fitted $J_\perp S\approx 44$ meV for $x\leq 0.21$, also much larger than the intralayer couplings. \cite{XLu2026Doping} The smaller $J_\perp$ for the compressively strained thin films seems to be consistent with their elongated $c$-axis lattice constants as compared to the bulk sample. \cite{Hwang2025,Chen2025}  Theoretically, the interlayer superexchange estimated from first-principles calculations of thin films is reduced by about 36\% and 24\% for single-stacked and double-stacked bilayer nickelates with respect to the pressurized bulk. \cite{DXYao2025film} 
\end{itemize}

\begin{itemize}[leftmargin=2.0em]
	\item\textit{Superconducting transition temperature.---}The bulk {\it R}$_{3}$Ni$_2$O$_7$ shows a transition temperature around $80$ K at high pressures, with a record high $T_c\approx 100$ K achieved recently by chemical substitution to reduce the interlayer distance. \cite{JJZhang2025,MWang2025,MWang2026Sm} Thin films at ambient pressure exhibit superconductivity at a much lower $T_c \sim 40$ K, \cite{Hwang2025,Chen2025} with the record $T_c\approx 63$ K achieved in high-quality (La,Pr)$_3$Ni$_2$O$_7$ thin films. \cite{ZYChen60K} Such a reduction of $T_c$ in thin films may be attributed to the reduced interlayer magnetic coupling associated with the elongated $c$-axis lattice constant. This was supported by recent observations that the $T_c$ of thin films  also increases under hydrostatic pressure. \cite{HHWen2026pressure,Tsukazaki2025Strain,Osada2026} In addition, a universal relationship has been found where the maximal $T_c$ decreases almost linearly with the $c$-axis lattice constant for both bulk and thin film samples, \cite{MWang2025} indicating a close relation between the interlayer coupling and superconductivity. 
\end{itemize}

\begin{itemize}[leftmargin=2.0em]
	\item\textit{$d_{z^2}$ metallization.---}In bulk La$_{3}$Ni$_2$O$_7$, the tilted Ni-O-Ni bonds at ambient pressure are straightened through a structural phase transition at high pressure, leading to an upward shift of the $d_{z^2}$ bonding band (self-doping) and the emergence of $\gamma$ hole pockets. \cite{MWang2023Nature,DXYao2023,Werner2023} Such a metallization of the $d_{z^2}$ bonding band was thought to play a crucial role in driving the superconductivity. \cite{MWang2023Nature,TXiang2015metalization,Ryee2025} This was supported by a recent study of tetragonal La$_3$Ni$_2$O$_7$ polycrystals, which suggests pressure-induced $d_{z^2}$ metallization for the superconductivity  without a structural transition. \cite{JingGuo2026}  For (La,Pr)$_3$Ni$_2$O$_7$ and (La,Sr)$_3$Ni$_2$O$_7$ thin films, ARPES measurements reported conflicting results with presence \cite{QKXue2025,ZYChen2026} or absence of $\gamma$ hole pockets, \cite{ZXShen2025,Nie2025ARPES} but all with nearly the same magnitude of $T_c$. However, it should be noted that the absence of $\gamma$ hole pockets does not necessarily mean that the $d_{z^2}$ orbital is fully localized, because the $\alpha-\beta$ band splitting of $d_{x^2-y^2}$ orbital can only arise from their hybridization with the $d_{z^2}$ quasiparticle band. \cite{Wang2026,Wang2026unified,Watanabe2026VMC} In addition, the compressive strain through LSAO substrate and oxygen annealing have proved to be crucial for the emergence of superconductivity, \cite{Nie2025PRL_crossover,Hwang2026halfdome,ZYChen2025SIT} both of which are found to enhance the $d_{z^2}$ metallization in recent RIXS and X-ray absorption (XAS) studies. \cite{KJZhou2026}   
\end{itemize}

\begin{itemize}[leftmargin=2.0em]
	\item\textit{Normal state properties.---}While the pressurized bulk La$_{3}$Ni$_2$O$_7$ shows perfect linear-in-$T$ resistivity above optimal $T_c$, \cite{HQYuan2024,Shen2025,JJZhang2025} most thin films exhibit FL transport in their normal states. \cite{Hwang2025,Chen2025,Hwang2025FL} Recent experimental studies on high-quality (La,Pr)$_3$Ni$_2$O$_7$ thin films \cite{ZYChen60K} and La$_2$LnNi$_2$O$_7$ (Ln = lanthanide) films grown on LSAO substrate \cite{Osada2026} found NFL normal-state resistivity $\Delta \rho\propto T^{\alpha}$, with the exponent $1\leq\alpha\leq 2$ correlated with $T_c$ ($\alpha\approx 1$ for samples with the highest $T_c$).  NFL behavior has also been reported in superconducting films grown on LAO substrate.  \cite{Hwang2025Reducing,Kumar2026} In addition, by tuning Sr doping \cite{Nie2025PRL_crossover,Nie2025Sr} or hydrostatic pressure, \cite{HHWen2026pressure} a crossover from metallic to weakly insulating behaviors has been observed recently, whose origin is still poorly understood. Moreover, several experiments reported the observation of pseudogap-like features above the superconducting phase. \cite{LYang2024,ZYChen2025,ZYChen2026Three}
\end{itemize}

\begin{itemize}[leftmargin=2.0em]
	\item\textit{Pairing symmetry.---}Scanning tunneling microscopy (STM) measurements of La$_2$PrNi$_2$O$_7$ thin films  support an  anisotropic $s^{\pm}$-wave gap symmetry. \cite{HHWen2025} Direct ARPES measurements of (La,Pr,Sm)$_3$Ni$_2$O$_7$ films suggest a weakly anisotropic $s$-wave gap on the $\beta$ sheet and an isotropic gap on the $\gamma$ sheet. \cite{ZYChen2025} The same group recently reported the observation of superconducting gaps on all $\alpha$, $\beta$ and $\gamma$ pockets, with a large gap ratio $2\Delta/k_\text{B}T_c\approx 8$  on the $\gamma$ surface. \cite{ZYChen2026Three} Ref. \cite{Nie2025ARPES} performed ARPES measurements on (La,Sr)$_3$Ni$_2$O$_7$ films and found gaps along the zone diagonal directions for both $\alpha$ and $\beta$ surfaces. Similar ARPES results have been reported recently on  La$_2$PrNi$_2$O$_7$/NdAlO$_3$ heterostructure. \cite{XHChen2026Thermo} Under high pressure, point-contact spectroscopy has been applied to  bulk bilayer nickelates to detect superconducting pairing symmetry, but the conclusion remains debated. \cite{HHWen2025Andreev,XLHuang2025Revealing,HkMao2025Direct}
\end{itemize}

The above experiments summarize the most important features of bilayer nickelates, which pose strong restrictions on theoretical descriptions. Considering the multiorbital electronic structure and electron interactions, one may consider the following bilayer two-orbital Hubbard model as a general starting point: \cite{DXYao2023,FYang2023,QHWang2023,JPHu2025PRL}
\begin{eqnarray}
	H&=&H_0+H_U, \label{eq:Hb} \\
	H_0&=&\sum_{ks}\Psi_{ks}^\dagger \mathcal{H}(k)\Psi_{ks},\notag \\
	H_U&=&\sum_{liss',a<b}\left(U'n_{lias}n_{libs'}+J_\text{H}d_{lias}^\dagger d_{libs}d_{libs'}^\dagger d_{lias'}\right)\notag \\
	&&+U\sum_{lia}n_{lia\uparrow}n_{lia\downarrow}+J_\text{P}\sum_{li,a\neq b}d_{lia\uparrow}^\dagger d_{lia\downarrow}^\dagger d_{lib\downarrow}d_{lib\uparrow}. \notag
\end{eqnarray}
Here $H_0$ is the tight-binding Hamiltonian obtained from DFT calculations, and $H_U$ is the multi-orbital Coulomb interaction. $\Psi_{ks}$ denotes the Fourier transform of the four-component vector $\Psi_{is}= (d_{1ixs},d_{1izs},d_{2ixs},d_{2izs})^{T}$, where $d_{lias}$ on layer $l=1,2$ annihilates a $d_{x^2-y^2}$ ($a=x$) or $d_{z^2}$ ($a=z$) electron at site $i$ with spin $s$. The tight-binding parameters are contained in the matrix $\mathcal{H}(k)$. \cite{DXYao2023} $U$, $U'$, $J_\text{H}$ and $J_\text{P}$ are the intraorbital and interorbital Coulomb repulsion, the Hund's coupling and the pair hopping interaction, respectively. The Kanamori relation $U'=U-2J_\text{H}$ and $J_\text{P}=J_\text{H}$ are often used. 

For moderate-to-strong electronic correlation relevant to bilayer nickelates, Eq. (\ref{eq:Hb}) is extremely difficult to solve, hence one must resort to various numerical methods or analytical approximations to extract useful information. An important simplification is to treat the two orbitals differently: the nearly half-filled $d_{z^2}$ orbital exhibits strong correlation and interlayer coupling, while the quarter-filled $d_{x^2-y^2}$ orbital is less correlated and forms an itinerant band. The two orbitals couple to each other through nearest-neighbor hybridization and local Hund's coupling, as described by the following effective model: \cite{YFYang2023}
\begin{eqnarray}
	H&=&-\sum_{lijs}t_{ij}c_{lis}^\dagger c_{ljs}-\sum_{lijs}(V_{ij}d_{lis}^\dagger c_{ljs}+H.c.) \notag \\
	&&+J_\text{H}\sum_{li}\mathbf{S}_{li}\cdot \mathbf{s}_{li}+H_\perp,
	\label{eq:HtJ} \\
	H_\perp&=&-t_{\perp}\sum_{is}(d_{1is}^\dagger d_{2is}+H.c.)+U\sum_{li}n_{li\uparrow}^d n_{li\downarrow}^d.\notag
\end{eqnarray} 
Here $c_{lis}^\dagger$ ($d_{lis}^\dagger$) creates a $d_{x^2-y^2}$ ($d_{z^2}$) electron at site $i$ and layer $l$, $n_{lis}^d=d_{lis}^\dagger d_{lis}$ is the $d_{z^2}$ occupation number per spin, $\mathbf{S}_{li}$ and $\mathbf{s}_{li}$ are the spin operators of $d_{z^2}$ and $d_{x^2-y^2}$ electrons, respectively. $t_{ij}$ and $t_\perp$ are the intralayer and interlayer hopping amplitudes of $d_{x^2-y^2}$ and $d_{z^2}$ electrons, $V_{i,i+x}=-V_{i,i+y}=V$ is the nearest-neighbor hybridization, and $U$ is the onsite Coulomb repulsion of $d_{z^2}$ orbital. The Coulomb repulsion of the $d_{x^2-y^2}$ orbital is not explicitly considered, but its renormalization effect has been absorbed into other parameters in Eq. (\ref{eq:HtJ}). \cite{YFYang2023} The term $H_\perp$ describes the atomic limit of the nearly half-filled $d_{z^2}$ orbital, which gives rise to an interlayer superexchange for $t_\perp \ll U$, 
\begin{equation}
	H_J=J\sum_{i}\mathbf{S}_{1i}\cdot \mathbf{S}_{2i}. \label{eq:HJ}
\end{equation} 
Strong correlation theories often take Eq. (\ref{eq:HJ}) as their starting point, where the $d_{z^2}$ spins form a valence bond state (VBS) consisting of interlayer spin singlets. This interlayer VBS may be viewed as the ``parent state" of the bilayer nickelates, which may lead to interlayer pairing superconductivity by considering its coupling to the $d_{x^2-y^2}$ metallic bands.

Note that more complicated $t$-$J$ models have been proposed and studied in the literature, which will not be covered in detail in this short review. For example, Ref. \cite{Kuroki2025tJ} derived a rather complicated $t$-$J$ model through Schrieffer-Wolff transformation from the bilayer two-orbital Hubbard model. Similar $t$-$J$ models (referred to as the type-II $t$-$J$ model in Ref. \cite{YHZhang2023}) are studied in Refs. \cite{YHZhang2023,FWang2024CPL,CJWu2025,YHZhang2025}.

\section{III. THEORIES OF INTERLAYER PAIRING SUPERCONDUCTIVITY}

In this section, we review important theoretical progress on interlayer pairing mechanisms of bilayer nickelates. Our focus will be on different aspects and various limits of the effective model, Eq. (\ref{eq:HtJ}). Emphasis will be placed on physical pictures and major theoretical predictions.  We will start from the atomic limit, i.e., the interlayer VBS formed by $d_{z^2}$ spins in the limit $J=\infty$, then introduce different pairing scenarios by considering different limiting situations.

\subsection{A. The atomic limit: interlayer valence bonds}

\begin{figure}[t]
	\begin{centering}
		\includegraphics[width=0.6\textwidth]{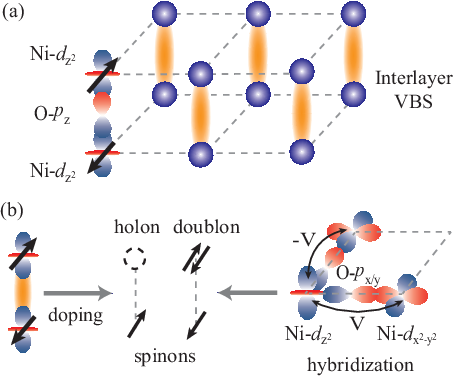}
		\par\end{centering}
	\caption{\textbf{Interlayer valence bonds.} (A) Schematic illustration of the $d_{z^2}$ interlayer valence bond state (VBS) in the atomic limit of bilayer nickelates. (B) Schematic diagram showing emergent $d_{z^2}$ holons, doublons and spinons induced by doping or interorbital hybridization. }
	\label{fig1}
\end{figure}

In the atomic limit, the nearest-neighbor hybridization vanishes due to the zero overlap between the $d_{z^2}$ orbital and the neighboring oxygen $p_{xy}$ orbitals.  The Hund's coupling is still present in this limit, but it will be temporarily neglected because there is only one $d_{x^2-y^2}$ electron on average for a single vertical Ni-O-Ni bond. Mathematically, this corresponds to the limit $V=J_\text{H}=0$  or equivalently $J=\infty$. In this case, the $d_{z^2}$ spins form a VBS consisting of independent interlayer spin singlets as schematically shown in Figure \ref{fig1}A: \cite{YFYang2023}
\begin{equation}
	\left|\text{VBS}\right\rangle=\prod_{i}\frac{1}{\sqrt{2}}\left(\left|\Uparrow\right\rangle_{1i}\left|\Downarrow\right\rangle_{2i}-\left|\Downarrow\right\rangle_{1i}\left|\Uparrow\right\rangle_{2i}\right), \label{eq:VBS}
\end{equation}   
where $\left|\sigma\right\rangle_{li}$ denotes the $d_{z^2}$ spin state with $\sigma=\Uparrow,\Downarrow$ on layer $l$. Note that such a featureless gapped Mott state is forbidden by the Lieb-Schultz-Mattis theorem in single-layer cuprate systems, but is allowed here in bilayer nickelates. \cite{YZYou2023,Hastings2004LSM} The gap towards the first excited state is given by a single singlet-to-triplet excitation (a ``triplon"), which costs an energy $\Delta_\text{VBS}=J$. In fact, Eq. (\ref{eq:HJ}) gives rise to a series of discrete energy levels $E_n=E_0+nJ$, where $n$ is the number of triplons and $E_0$ is the ground state energy. Note that these triplons can be produced either through thermal fluctuations or quantum fluctuations.  The latter occurs when the $d_{z^2}$ spin is coupled to the $d_{x^2-y^2}$ spin through either Hund's coupling or Kondo coupling (a second-order effect of the hybridization), since such terms do not commute with Eq. (\ref{eq:HJ}). The triplons may mediate an effective attractive interaction between $d_{x^2-y^2}$ electrons and induce superconductivity, \cite{Khaliullin2026,HLiu2026triplon} similar to the phonon-mediated BCS superconductivity.

One important question is how $d_{z^2}$ doping may affect the VBS. Because the singlets at different sites are completely disentangled in the atomic limit, the effect of doping is simply to reduce the number of singlet bonds and to introduce local fractionalized particles --- spinons (unpaired $d_{z^2}$ spins), holons ($d_{z^2}$ holes) and doublons ($d_{z^2}$ doubly occupied states). The hybridization-induced self-doping has a similar effect, as schematically shown in Figure \ref{fig1}B. These fractionalized objects can interact with $d_{x^2-y^2}$ electrons once their couplings are considered, rationalizing a slave-particle description of bilayer nickelates as will be discussed later. \cite{Wang2025,Wang2026,Wang2026unified} If the average hole concentration per layer is $\delta_d$ (corresponding to a $d_{z^2}$ occupation $n_d=1-\delta_d$), the probability of two $d_{z^2}$ spins at different layers to form a spin singlet bond is $(1-|\delta_d|)^2$,  assuming the holes (or doublons) equally occupy the two layers. This leads to an effective interlayer magnetic coupling $J_\text{eff}=J (1-|\delta_d|)^2$. A similar form of effective magnetic coupling under doping has also been found in cuprates. \cite{Devereaux2013} 

\subsection{B. Interlayer pairing mechanisms}

Starting from the atomic-limit interlayer VBS, we now introduce its coupling to the itinerant $d_{x^2-y^2}$ orbital, which allows the system to build superconductivity. Theoretically, both the interorbital hybridization \cite{YFYang2023,QQin2023,QQin2024,YFYang2024,YFYang2025CPL,Wang2025,GMZhang2023,Wang2026unified,WWu2025} and Hund's rule coupling \cite{CJWu2024,CJWu2024PRL,CJWu2025,FYang2025Pairing,FYang2026Unified} have been proposed to induce interlayer pairing superconductivity, which correspond to different limiting considerations. In the limit of fully localized $d_{z^2}$ electrons, the Hund's coupling and the hybridization-induced Kondo coupling may together trigger triplon-mediated superconductivity. \cite{Khaliullin2026,HLiu2026triplon} \\

\noindent\textbf{\textit{$J_\text{H}=0$ limit}}

The consideration of the $J_\text{H}=0$ limit is justified by the following reasons: i) The Hund's coupling between the localized $d_{z^2}$ orbital and the itinerant $d_{x^2-y^2}$ electrons acts as a ferromagnetic Kondo coupling, which flows to the weak-coupling limit under the renormalization group for a single magnetic impurity. \cite{Anderson1970} For the lattice, a previous study of the two-orbital $t$-$J_H$-$J$ model suggests a similar weakening of the Hund's coupling at low temperature, with the normal state remaining a Fermi liquid for arbitrary $J_H$. \cite{Wang2026} ii) The $d_{z^2}$ moment will be suppressed by doping or hybridization, leading to a reduced effective $J_\text{H}$ especially at high pressure; iii) According to DFT+DMFT calculations, the major effect of Hund's coupling is to enhance the quasiparticle effective mass and lead to flat $d_{z^2}$ quasiparticle bands near the Fermi energy. \cite{YYCao2024} Therefore, for simplicity, it is instructive to absorb the Hund's coupling by a renormalization of the other parameters and construct a minimal $t$-$V$-$J$ model for the superconductivity, which corresponds to the $J_\text{H}=0$ limit of Eq. (\ref{eq:HtJ}). Based on this minimal model, Ref. \cite{YFYang2023} proposed a two-component theory for the bilayer nickelate superconductivity, following an earlier proposal by Kivelson for high-$T_c$ cuprates. \cite{Kivelson2002}  The basic idea is that the $d_{z^2}$ electrons form local interlayer singlet pairs, and their hybridization with the metallic $d_{x^2-y^2}$ electrons induces global superconducting phase coherence.

\begin{figure}[t]
	\begin{centering}
		\includegraphics[width=0.6\textwidth]{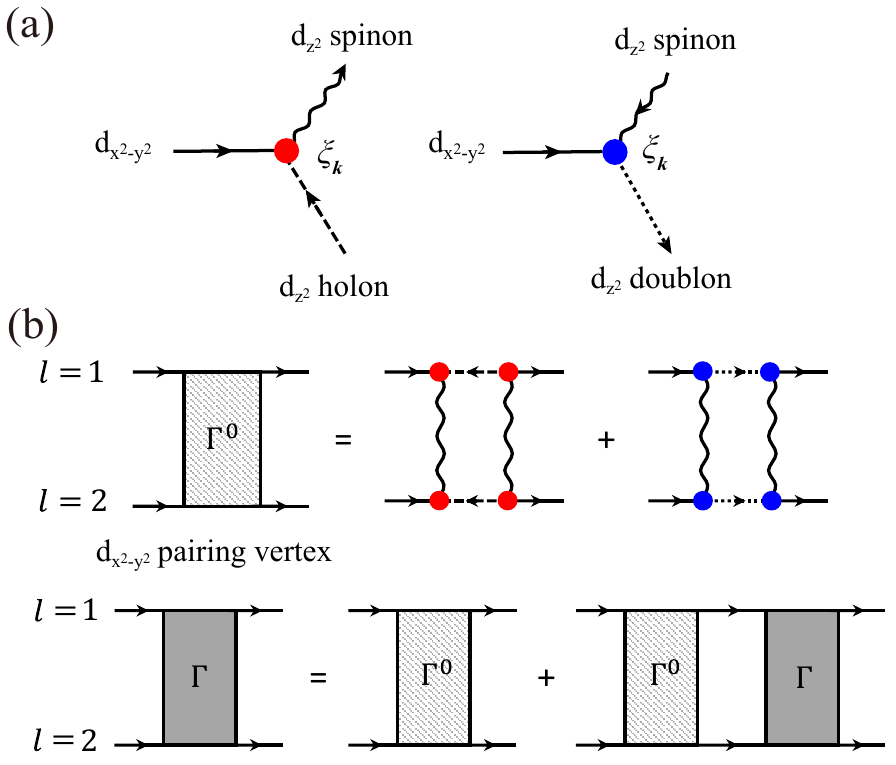}
		\par\end{centering}
	\caption{\textbf{Hybridization-driven superconductivity.} (A) Electron-spinon-holon/doublon scattering vertices induced by the nearest-neighbor interorbital hybridization with the factor $\xi_{\bf k}=\cos k_x-\cos k_y$. (B) The bare ($\Gamma_0$) and full ($\Gamma$) interlayer pairing vertex of $d_{x^2-y^2}$ electrons. $\Gamma_0$ is constructed from $d_{z^2}$ spinon and holon (doublon) propagators, while $\Gamma$ is obtained from $\Gamma_0$ and $d_{x^2-y^2}$ propagators through the Bethe-Salpeter equation. }
	\label{fig2}
\end{figure}

In this picture, the $d_{z^2}$ interlayer superexchange provides a high pairing energy scale, which is represented by its mean-field pairing amplitude $\Delta_d$ through the decomposition $J \sum_{i}\mathbf{S}_{1i}\cdot \mathbf{S}_{2i}\rightarrow -\Delta_d^* \psi_{i}^d$, where $\psi_i^d=\frac{1}{\sqrt{2}}\sum_{ss'}d_{1is}(i\sigma_{ss'}^y)d_{2is'}$ is the $d_{z^2}$ interlayer singlet pair. A second-order perturbation with $V$ then yields the pairing term for the $d_{x^2-y^2}$ orbital, 
\begin{equation}
	H_\Delta=-\frac{1}{\sqrt{2}}\sum_{{\bf k}ss'}\Delta_c^*({\bf k})c_{1\mathbf{k}s}(i\sigma_{ss'}^y)c_{2-{\bf k}s'}+H.c. \label{eq:HDelta}
\end{equation}
where $\Delta_c(\mathbf{k})=\tilde{\Delta}_c(\cos k_x-\cos k_y)^2$ with $\tilde{\Delta}_c\sim\delta_d V^2/\Delta_d$. \cite{YFYang2023} The factor $\xi_{\bf k}=\cos k_x-\cos k_y$ comes from the Fourier transformation of the anisotropic hybridization $V_{i,i+x}=-V_{i,i+y}=V$. Therefore, it predicts an isotropic $s$-wave pairing on the $d_{z^2}$ Fermi surface, and an anisotropic extended $s$-wave pairing with possible nodes or gap minima along the zone diagonal on the $d_{x^2-y^2}$ Fermi surfaces. Considering the bonding-antibonding splitting, the gap functions on the bonding $\alpha$ and $\gamma$ bands have the same sign, while the antibonding $\beta$ band has the opposite sign.  Such an anisotropic $s^{\pm}$-wave pairing is qualitatively consistent with current experimental observations on thin films.\cite{HHWen2025,ZYChen2025} Note that the gap nodes on the $\alpha$ and $\beta$ bands are not protected by symmetry, so they may become gap minima in real materials due to many other effects. For example, the $d_{x^2-y^2}$ electrons may form a weak isotropic interlayer pairing through the small vertical hopping or Hund's coupling, or in-plane extended $s$-wave pairing through the small intralayer superexchange, or simply due to disorder. 

Theoretically, the above pairing symmetry was confirmed by numerical auxiliary-field Monte Carlo simulations \cite{QQin2023,QQin2024} and dynamic Schwinger boson calculations based on the $t$-$V$-$J$ model.\cite{Wang2025,Wang2026unified} In the Schwinger boson approach, \cite{Wang2026unified} the $d_{z^2}$ operator is represented as $d_{lis}=b_{lis}\chi_{li}^\dagger+sb_{li-s}^\dagger \zeta_{li}$ by using the bosonic spinons $b_{lis}$, fermionic holons $\chi_{li}$, and fermionic doublons $\zeta_{li}$. The hybridization term then becomes three-particle scattering vertices illustrated in Figure \ref{fig2}A, which gives rise to self-energies for all slave particles and $d_{x^2-y^2}$ electrons and hence their various anomalous normal state properties. Solving the Bethe-Salpeter equation shown in Figure \ref{fig2}B gives the interlayer pairing vertex $\Gamma_{ss'}(\mathbf{k},i\omega_n;\mathbf{k}',i\omega_{n'})=ss'\xi_{\bf k}^2\xi_{\mathbf{k}'}^2\tilde{\Gamma}(i\omega_n,i\omega_{n'})$ for $d_{x^2-y^2}$ electrons, indicating anisotropic $s^\pm$-wave pairing gaps for $\alpha$ and $\beta$ Fermi surfaces. The interlayer pairing gap of $d_{z^2}$ electrons is local on the $ab$ plane, $\sum_s\langle sd_{1is}d_{2j\bar{s}} \rangle \propto \delta_{ij}$, explaining the isotropic $s$-wave gap on the $\gamma$ pocket.

The hybridization-driven superconductivity has been supported by other theoretical investigations from different groups. \cite{GMZhang2023,WWu2025,WWu2025Intertwined,Watanabe2026VMC,GSu2025} For example, Ref. \cite{WWu2025} studied the bilayer two-orbital Hubbard model using the cluster dynamical mean-field theory (CDMFT), and found an $s^{\pm}$-wave pairing driven by the hybridization and $d_{z^2}$ vertical hopping. Ref. \cite{Watanabe2026VMC} studied the bilayer two-orbital Hubbard model through the variational Monte Carlo method, and found a hierarchical pairing structure qualitatively consistent with the two-component picture. Density-matrix renormalization group (DMRG) calculations on quasi-one-dimensional systems also confirmed the existence of hybridization-driven superconductivity. \cite{GMZhang2023,GSu2025}\\

\begin{figure}[t]
	\begin{centering}
		\includegraphics[width=0.55\textwidth]{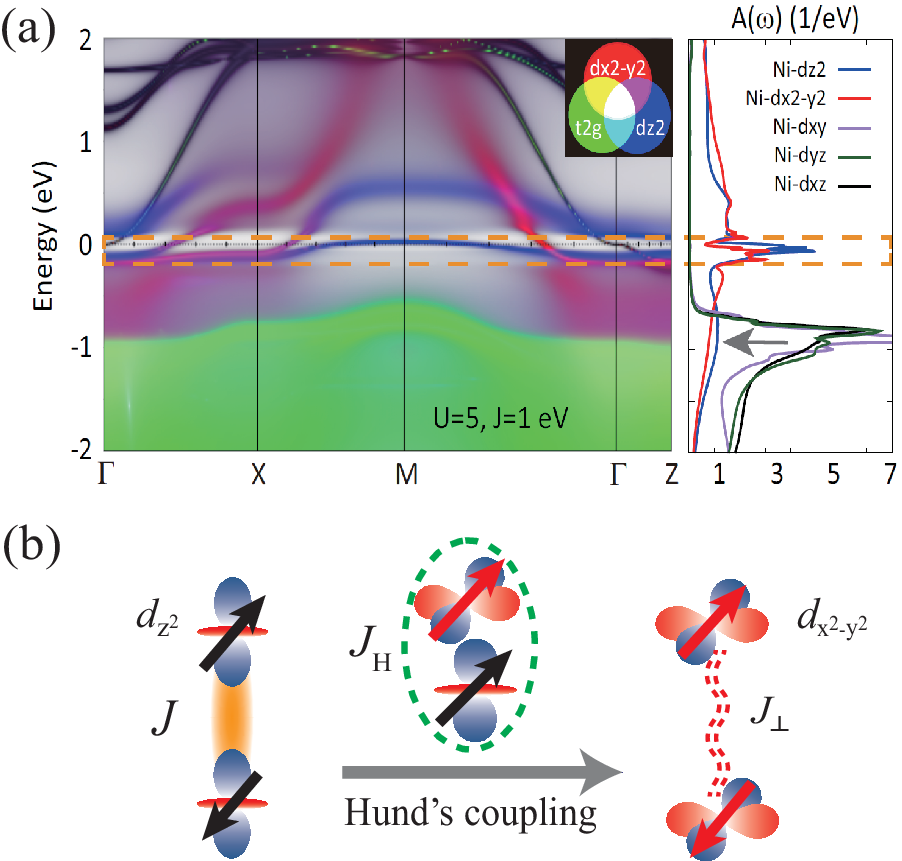}
		\par\end{centering}
	\caption{\textbf{Effects of the Hund's rule coupling.} (A) Orbital-projected spectral function and densities of states of Ni-$d$ electrons calculated using DFT+DMFT for $U=5$ eV and Hund's coupling $J_\text{H}=1$ eV, showing emergent flat band and orbital-dependent renormalization promoted by the Hund's coupling. Adapted from Ref. \cite{YYCao2024}. (B) Schematic diagram showing the effective interlayer $J_\perp$ between $d_{x^2-y^2}$ orbitals transmitted from $d_{z^2}$ orbitals through the Hund's coupling. }
	\label{fig3}
\end{figure}

\noindent\textbf{\textit{$J_\text{H}=\infty$ limit}}

In addition to the band renormalization effect shown in Figure \ref{fig3}A from DFT+DMFT calculations, \cite{YYCao2024} the Hund's coupling was also proposed to drive interlayer pairing superconductivity. \cite{CJWu2024PRL} The basic idea is that the strong Hund's rule coupling leads to parallel spin alignment of $d_{z^2}$ and $d_{x^2-y^2}$ electrons at the same Ni atom. Therefore, the interlayer magnetic coupling between $d_{z^2}$ spins can be effectively transmitted to the $d_{x^2-y^2}$ orbital, leading to interlayer pairing superconductivity of $d_{x^2-y^2}$ electrons (see Figure \ref{fig3}B for a schematic illustration). This mechanism is naturally associated with the limit $J_\text{H}=\infty$, in which case the $d_{z^2}$ and $d_{x^2-y^2}$ electrons on the same Ni atom form an effective $S=1$ spin. Together with the interlayer spin coupling, the Ni-O-Ni bonds now form another spin-1 singlet valence bond state other than the spin-$\frac{1}{2}$ VBS introduced above, and the hole doping upon this mixed VBS then causes superconductivity. \cite{CJWu2025,YHZhang2023,YHZhang2025} This mechanism was initially studied based on a single-orbital bilayer $t$-$J_\perp$-$J_\parallel$ model of the $d_{x^2-y^2}$ orbital, obtained by integrating out the $d_{z^2}$ orbital in the large $J_\text{H}$ limit. \cite{CJWu2024PRL} Here $J_\perp$ and $J_\parallel$ denote the interlayer and intralayer magnetic coupling of $d_{x^2-y^2}$ electrons, respectively. For large ratios of $J_\perp/J_\parallel$ and  electron fillings relevant to bilayer nickelates, slave-boson mean field theory predicts an interlayer $s$-wave pairing symmetry. \cite{CJWu2024PRL,CJWu2024CPL} Due to the quarter filling of $d_{x^2-y^2}$ orbital, the $T_c$ of Hund-driven superconductivity is determined by the spinon pairing temperature similar to the overdoped cuprates, \cite{CJWu2024PRL} which is different from the hybridization scenario where $T_c$ is mainly determined by the phase coherence temperature. \cite{YFYang2023,QQin2023} The single-orbital model was later studied numerically by DMRG \cite{Bohrdt2024DMRG,FYang2026Unified,ZYWeng2024} and infinite projected entangled-pair state (iPEPS) method. \cite{GSu2024} Away from half-filling, the $d_{z^2}$ orbital cannot be simply integrated out, hence it requires two-orbital $t$-$J$ models to correctly describe the low-energy physics. Such two-orbital $t$-$J$ models have been studied via parton mean-field theory, \cite{YHZhang2023, CJWu2024,CJWu2025,FYang2025Pairing} multiband Gutzwiller approximation \cite{FWang2024CPL} and DMRG method \cite{CJWu2025} in the large-$J_\text{H}$ limit, with the results qualitatively consistent with the single-band $t$-$J_\perp$-$J_\parallel$ model.\\

\noindent\textbf{\textit{Finite-$J_\text{H}$}}

The large-$J_\text{H}$ calculations neglected important quantum fluctuations of the Hund interaction, which should play a role in real materials. Therefore, it is more realistic to study the bilayer two-orbital model Eq. (\ref{eq:HtJ}) with finite $J_\text{H}$. To focus on the Hund-driven superconductivity, one may take $V=0$ to exclude the hybridization mechanism. Such a bilayer two-orbital $t$-$J_\text{H}$-$J$ model was studied using the dynamic Schwinger boson approach in Ref. \cite{Wang2026}. By comparing with the results of the $t$-$V$-$J$ model under the same approach, \cite{Wang2025} Ref. \cite{Wang2026} found that the Hund mechanism gives several different predictions: 1) The pairing symmetry is a fully isotropic $s^{\pm}$-wave due to the locality of the Hund's coupling. This conclusion does not change qualitatively when small intralayer superexchange couplings are included. \cite{CJWu2025,FYang2025Pairing} 2) The normal state is generally a Fermi liquid. This is because the Hund's coupling between the localized $d_{z^2}$ orbital and the itinerant $d_{x^2-y^2}$ electrons acts as a ferromagnetic Kondo coupling, which is expected to be weakened under renormalization group flow, resulting in a weakly correlated $d_{x^2-y^2}$ Fermi liquid. 3) $T_c$ decreases monotonically as the $d_{z^2}$ hole doping increases, suggesting that $d_{z^2}$ metallization is detrimental to the Hund-driven superconductivity. 4) The maximal ratio $T_c/J$ for the Hund scenario is smaller than that of the hybridization scenario based on mean-field calculations. These are summarized in Table \ref{tab1}.

\begin{table*}
	\centering
	\caption{Comparison between the Schwinger boson results of the $t$-$V$-$J$ model (hybridization scenario) in the large-$U$ limit\cite{Wang2025} and the $t$-$J_H$-$J$ model (Hund scenario). \cite{Wang2026} A correction factor of 0.3 is introduced for the $T_c^\text{max}$ to tentatively account for the effect of phase fluctuations based on previous comparison with Monte Carlo simulations. \cite{QQin2023} SC stands for superconductivity. Adapted from Ref. \cite{Wang2026}.}
	\label{tab1}
	\begin{tabular}{lcccc}
		\hline \hline
		& $\quad$ & $t$-$V$-$J$ &  $\quad$ & $t$-$J_H$-$J$   \\
		\hline
		Pairing symmetry &  $\quad$ & Anisotropic $s^\pm$-wave &  $\quad$ & Isotropic $s^\pm$-wave \\ 
		$0.3T_c^\text{max}/J$ & $\quad$  & $\approx 0.05$ & $\quad$  & $\approx 0.03$   \\ 
		$d_{z^2}$ metallization &  $\quad$ & Crucial for SC & $\quad$  & Harmful for SC \\
		Normal state & $\quad$  &NFL around optimal $T_c$ & $\quad$  & FL for all parameters\\
		\hline \hline
	\end{tabular}
\end{table*}

Some authors investigated the effects of hybridization and Hund's coupling simultaneously, for example, through DMRG calculations of $t$-$J$ models \cite{Kuroki2024DMRG,GSu2025} or CDMFT calculations of the bilayer two-orbital Hubbard model. \cite{WWu2025Intertwined} Both hybridization-driven and Hund-driven superconducting states have been found in these studies, and become dominant in different parameter regions. \cite{WWu2025Intertwined,GSu2025} For a more detailed review of the Hund mechanism, readers may refer to Ref. \cite{CJWu2026review}. \\

\noindent\textbf{\textit{$d_{z^2}$ local-moment limit}}

In the limit of completely localized $d_{z^2}$ spins, the triplon excitations from the interlayer VBS may mediate superconductivity for $d_{x^2-y^2}$ electrons through both Hund and Kondo couplings. \cite{Khaliullin2026,HLiu2026triplon} By treating these spin interactions using a second-order perturbation theory, Ref. \cite{HLiu2026triplon} obtained a pairing interaction for $d_{x^2-y^2}$ bands, similar to the phonon-mediated BCS superconductivity. The resulting pairing symmetry is anisotropic $s^{\pm}$-wave with gap minima along the Brillouin zone diagonal for both $\alpha$ and $\beta$ Fermi sheets. Roughly speaking, this mechanism combines the above-mentioned two ideas (hybridization and Hund) in the limit of localized $d_{z^2}$ spins. However, whether such a local picture can lead to high $T_c$ requires further investigations. Another issue of this picture is that it cannot explain the large superconducting gap on the $d_{z^2}$-dominated $\gamma$ Fermi surface observed in experiment. \cite{ZYChen2025,ZYChen2026Three}

\begin{figure*}[t]
	\begin{centering}
		\includegraphics[width=0.65\textwidth]{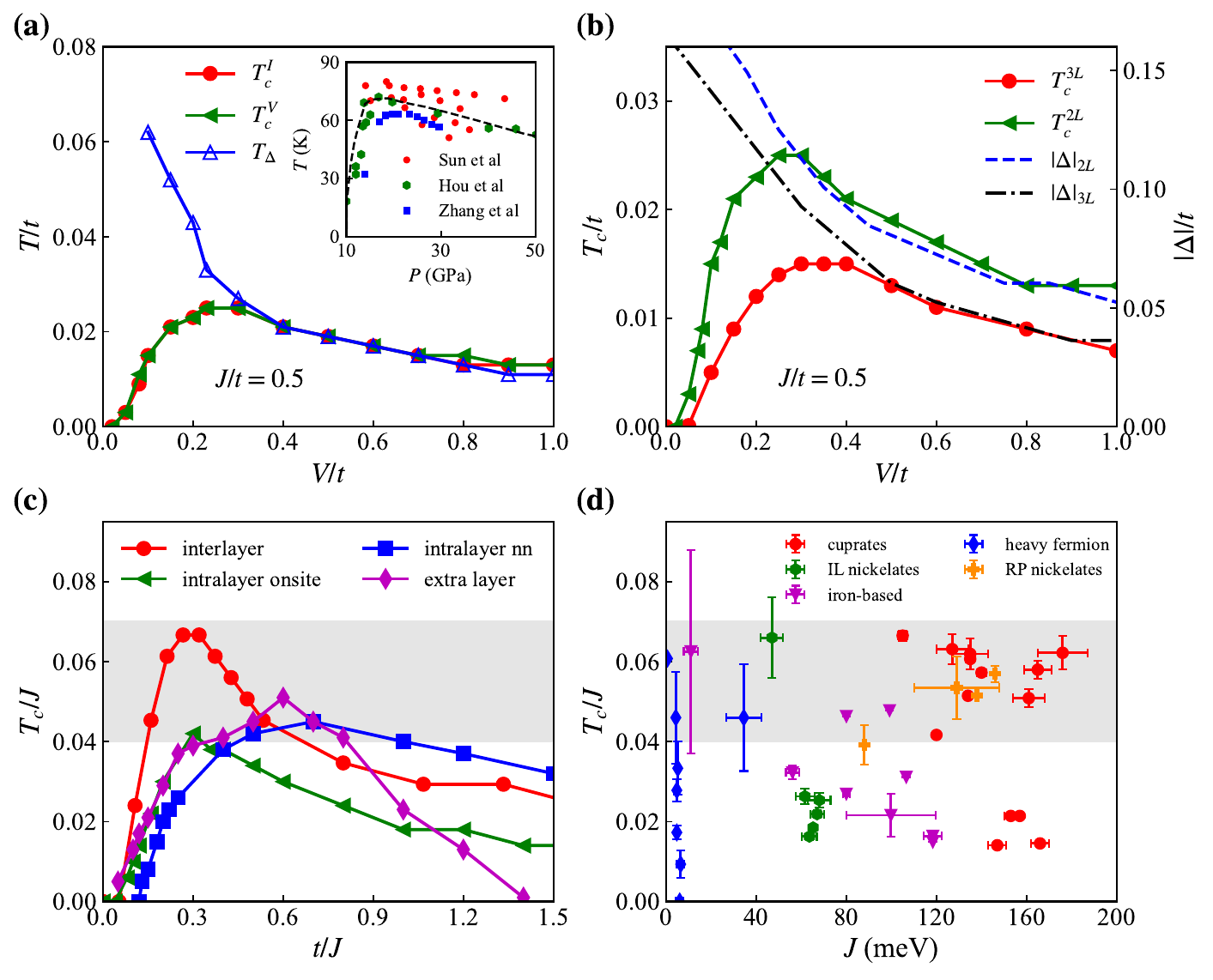}
		\par\end{centering}
	\caption{\textbf{Evolution of the superconducting transition temperature.} (A) $T_c$ from superconducting phase coherence as functions of the hybridization parameter $V$ derived from the mutual information ($T_c^I$) and the change of vortex-antivortex numbers ($T_c^V$), calculated from Monte Carlo simulations of the bilayer $t$-$V$-$J$ model with $J/t=0.5$. $T_\Delta$ is the $d_{z^2}$ local pairing temperature. The $d_{x^2-y^2}$ nearest-neighbor hopping $t$ is set as the energy unit. Adapted from Ref. \cite{QQin2023}. (B) Comparison of $T_c$ and the pairing amplitude $|\Delta|$ (the most probable value obtained from Monte Carlo simulations at low temperature $T/t=0.002$) for bilayer ($2L$) and trilayer ($3L$) $t$-$V$-$J$ models with $J/t=0.5$. Adapted from Ref. \cite{QQin2024}. (C) $T_c/J$ as functions of $t/J$ for different $t$-$J$ type effective models with interlayer, intralayer onsite, or intralayer nn (nearest-neighbor) pairing interactions of unconventional superconductivity obtained by Monte Carlo simulations. The nearest-neighbor attractive charge density interaction is also included for the superexchange mechanism. For the intralayer nearest-neighbor pairing, introducing an extra layer of conduction electrons with nearest-neighbor interlayer hopping (simulating the apical oxygens) is found to enhance the maximum $T_c/J$. (D) Collection of experimental $T_c/J$ ratios for different families of unconventional superconductors (see Ref. \cite{QQin2025} and references therein. Additional data for RP and infinite-layer (IL) nickelates are extracted from Refs. \cite{DLFeng2024,MWang2024,HLuo2026spin,XLu2026Doping,MWang2023Nature,JGCheng2024Nature,MWang2025} and Refs. \cite{JGCheng2022Pressure,Hwang2023Linear,Fan2024Capping,XJZhou2024Magnetic,Ariando2025Bulk,QSWang2026Persistent,DFLi2026}, respectively.) (C) and (D) are adapted from Ref. \cite{QQin2025}. }
	\label{fig4}
\end{figure*}

\subsection{C. Superconducting transition temperature}

In this subsection, we discuss the superconducting transition temperature and its dependence on different parameters, including the interlayer superexchange $J$ and the $d_{z^2}$ doping level. We will focus on major theoretical predictions of the hybridization mechanism and their comparisons with experimental observations. \\

\noindent\textbf{\textit{$T_c$ versus $V$ and $J$}}

In the two-component theory of bilayer nickelates, the $d_{z^2}$ interlayer valence bond amplitude $\Delta_d\sim J$ gives the pairing energy scale, while the true $T_c$ is associated with the two-dimensional Kosterlitz-Thouless transition and determined by the $d_{x^2-y^2}$ phase stiffness. \cite{YFYang2023} Starting from a large $J$ and a half-filled $d_{z^2}$ orbital, Ref. \cite{YFYang2023} gives a rough estimate $T_c\sim t(\delta_d V^2/J t)^{2/3}$, where $t$ is the $d_{x^2-y^2}$ in-plane hopping amplitude and $\delta_d$ is the $d_{z^2}$ hole doping. However, because of the competition between hybridization and the $d_{z^2}$ local pairing field, $T_c$ typically exhibits a nonmonotonic variation upon tuning $V$ or $J$. This was indeed confirmed by static auxiliary field Monte Carlo simulations of the $t$-$V$-$J$ model. \cite{QQin2023} Figure \ref{fig4}A shows the simulated $T_c$ as a function of $V$ for fixed $J/t=0.5$. When $V$ is small, the $d_{z^2}$ electrons form strong local singlet pairs with a large $T_\Delta$, but $T_c$ is much lower due to the difficulty of establishing long-range phase coherence. The region $T_c<T<T_\Delta$ corresponds to preformed pairs and may exhibit pseudogap phenomenon, as will be discussed later in more detail. As $V$ increases, $T_\Delta$ decreases monotonically due to the suppressed $d_{z^2}$ pairing field, and $T_c$ tracks $T_\Delta$  closely for large $V$, leading to a maximal $T_c$ at optimal hybridization.

To see whether the magnetic coupling provides the pairing energy scale, it is more relevant to compute the maximal ratio between $T_c$ and $J$, and make comparisons with experiments. Refs. \cite{QQin2023,QQin2024} computed this ratio for bilayer and trilayer $t$-$V$-$J$ models respectively, and found $T_c/J\sim 0.04-0.05$ for bilayer nickelates and $0.02-0.03$ for trilayer nickelates (see Figure \ref{fig4}B). If one takes the interlayer $J$ of bilayer and trilayer nickelates to be approximately 150 meV \cite{DLFeng2024} and 110 meV \cite{MWang2026Distinct}, respectively, from RIXS measurements (assuming the spin size $S\approx 1/2$), these ratios immediately predict a maximal $T_c$ of about $70$--$90$ K for bilayer nickelates and $25$--$40$ K for trilayer nickelates, both consistent with experiments. \cite{MWang2023Nature,JZhao2024} The reduced $T_c/J$ in trilayer nickelates is caused by the intrinsic frustration between interlayer singlet pairs formed by the two outer-layer Ni-3$d$ electrons with the same inner-layer $d$ electron, which is unique for interlayer pairing. \cite{QQin2024,YFYang2024} Remarkably, by comparing Monte Carlo simulations with experiments, Ref. \cite{QQin2025} proposed a practical maximal ratio $T_c/J\approx 0.04-0.07$ for all existing unconventional superconductor families (see Figure \ref{fig4}C-D), suggesting that the highest $T_c$ of unconventional superconductors is limited by the magnitude of the pairing strength $J$ rather than specific pairing mechanisms.    \\

\noindent\textbf{\textit{$T_c$ versus doping}}

Investigation of the Ni-3$d$ doping effect is important both theoretically and experimentally, since it helps to distinguish different pairing mechanisms and to resolve the contradictory observations regarding the presence/absence of $\gamma$ hole pockets in thin film bilayer nickelates. \cite{QKXue2025,ZYChen2026,ZXShen2025,Nie2025ARPES} From the strong correlation point of view, the effect of $d_{z^2}$ doping is more important than that of the $d_{x^2-y^2}$ orbital, because it determines the $d_{z^2}$ metallization, which is crucial for hybridization-driven superconductivity but detrimental for Hund-driven superconductivity. On the other hand, the major effect of $d_{x^2-y^2}$ doping is to vary the Fermi surface shape of $\alpha$ and $\beta$ bands along with their density of states (DOS), which plays a key role in weak correlation theories.           

In terms of the sensitivity of the transition temperature to $d_{z^2}$ doping, there seems to be a significant difference between the bulk and thin film bilayer nickelates. The bulk is non-superconducting at ambient pressure where the $\gamma$ band is located below the Fermi energy, while thin films show superconductivity in samples with or without the $\gamma$ Fermi surface. A unified theoretical explanation was recently proposed based on the two-component theory. \cite{Wang2026unified} The key to this issue lies in the different strength of interlayer magnetic coupling in bulk and thin film samples: bulk bilayer nickelates have a larger $J$ (or equivalently a larger binding energy for the $d_{z^2}$ valence bonds), while thin films have a smaller $J$, hence more weakly bound $d_{z^2}$ valence bonds, due to the larger interlayer distance induced by the compressive strain. While superconductivity requires the hybridization to induce phase coherence, a large $J$ strongly suppresses the hybridization at half filling, leading to nearly decoupled $d_{z^2}$ VBS and $d_{x^2-y^2}$ Fermi liquid. In this case, a minimal $d_{z^2}$ doping is required to delocalize the VBS and promote the hybridization-driven superconductivity. This explains why the bulk bilayer nickelates show superconductivity only in the presence of $\gamma$ pockets. On the other hand, the smaller $J$ in thin film nickelates already allows the hybridization to take place at half filling and induce superconductivity even without the $\gamma$ pockets. 

\begin{figure}[t]
	\begin{centering}
		\includegraphics[width=0.6\textwidth]{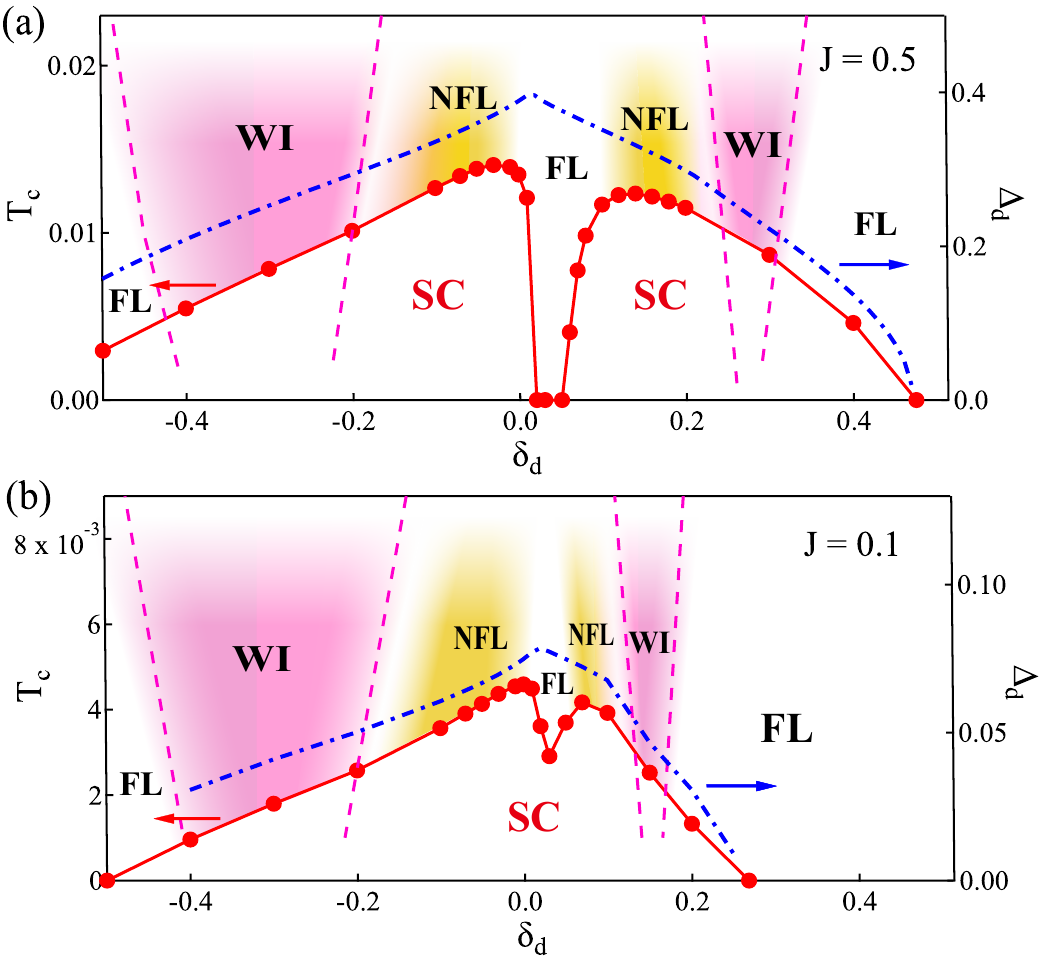}
		\par\end{centering}
	\caption{\textbf{Superconducting and normal state phase diagrams predicted by the two-component theory.} (A) $T_c$ as a function of $d_{z^2}$ doping $\delta_d$ for interlayer coupling $J=0.5$ at $V=0.5$ and $U=7$, obtained from dynamic Schwinger boson calculations of the $t$-$V$-$J$ model. For better comparison with experiment, a correction factor 0.3 for $T_c$ is used to tentatively account for phase fluctuations based on previous comparison with Monte Carlo simulations. \cite{QQin2023,Wang2026} The blue dash-dotted line is the $d_{z^2}$ valence bond amplitude $\Delta_d$ at low temperature $T=0.004$. The colors represent different normal states  including the Fermi liquid (FL), non-Fermi-liquid (NFL) and weakly insulating (WI) behaviors. The dashed lines mark crossovers between different regions. (B) Similar plot as (A), but for a smaller $J=0.1$. Adapted from Ref. \cite{Wang2026unified}.}
	\label{fig:doping}
\end{figure}

Theoretically, Figure \ref{fig:doping} shows $T_c$ as a function of $d_{z^2}$ doping $\delta_d$ for different $J$ calculated using the dynamic Schwinger boson approach for the $t$-$V$-$J$ model with a finite $U$. \cite{Wang2026unified} For large $J=0.5$, there are two superconducting domes located on both the electron doping ($\delta_d<0$) and hole doping ($\delta_d>0$) sides, separated by the non-superconducting VBS around half filling. While the highly asymmetric dome under hole doping is consistent with the superconducting phase in pressurized bulk La$_3$Ni$_2$O$_7$, the dome under electron doping has not yet been observed although it has been speculated. \cite{Hwang2026halfdome,MJiang2026EleDope,LiangSi2026,WeiWu2026Electron} The gradual suppression of $T_c$ at large hole (electron) doping is due to the reduced $d_{z^2}$ pairing amplitude $\Delta_d$. Considering the suppression of effective superexchange by doping, $T_c$ will decrease more rapidly at heavy doping levels. For small $J=0.1$, the two domes merge into a single one with a lower $T_c$, hence superconductivity survives even at half filling, consistent with experimental observations on thin films. \cite{QKXue2025,ZYChen2026,ZXShen2025,Nie2025ARPES}  In addition, similar nonmonotonic evolution of $T_c$ with $d_{z^2}$ hole doping for small $J$ has been observed experimentally in Sr-doped \cite{Nie2025Sr,Nie2025PRL_crossover} and pressurized thin film bilayer nickelates, \cite{HHWen2026pressure} where the pressure has been demonstrated to shift the $\gamma$ band upward and increase the $d_{z^2}$ hole doping. \cite{HHWen2026pressure} A recent study on heavily Sr-doped La$_{3-x}$Sr$_x$Ni$_2$O$_{7-\delta}$ shows that the holes introduced by Sr doping predominantly occupy the $d_{z^2}$ orbital, reducing its occupation to nearly half ($\delta_d\approx 0.5$) for $x=1$. \cite{HHWen2026Heavily} Assuming a linear $x\sim \delta_d$ relation, Fig. \ref{fig:doping} predicts the suppression of superconductivity for $x\gtrsim 1$ ($\delta_d\gtrsim 0.5$) at large $J$, and for $x\gtrsim 0.4$ ($\delta_d\gtrsim 0.2$) at small $J$, both in agreement with experimental observations. \cite{HHWen2026Heavily,Nie2025Sr} Near half-filling, the theory also predicts a maximal $T_c$ of about 50 K for moderate $U$ if only doping or the interlayer distance ($J$) is tuned. \cite{Wang2026unified}

In contrast to the hybridization scenario, the Hund scenario predicts a different $T_c$ dependence on $d_{z^2}$ doping. Since the Hund's rule coupling is a spin interaction, it is naturally weakened upon $d_{z^2}$ doping, which reduces the effective moment. Therefore, the $T_c$ of Hund-driven superconductivity  decreases rapidly as $d_{z^2}$ hole doping increases, as revealed by dynamic Schwinger boson studies of the $t$-$J_\text{H}$-$J$ model. \cite{Wang2026}  Similar conclusions have been obtained by DMRG \cite{GSu2025} and slave-boson mean field calculations \cite{CJWu2024,CJWu2025} of two-orbital $t$-$J$ models. As for the effect of $d_{x^2-y^2}$ doping, both slave-boson mean field theory and DMRG calculations of the single-orbital $t$-$J_\perp$-$J_\parallel$ model predict that $T_c$ decreases with  hole doping but increases with electron doping, as expected from the original quarter filling of the $d_{x^2-y^2}$ orbital. \cite{FYang2026Unified}   \\

\subsection{D. Normal state properties}

For unconventional superconductors, the nontrivial normal state is thought to be closely related to the superconducting pairing mechanism. It is thus important for  theories of bilayer nickelates to explain both the superconducting and normal state properties. Based on the two-component picture, Refs. 
\cite{Wang2025,Wang2026unified} have successfully explained the NFL and FL normal states in both pressurized bulk and thin film bilayer nickelates. Moreover, this theory predicts a continuous evolution from metallic to weakly insulating normal state behaviors with increasing $d_{z^2}$ doping, consistent with recent experiments on Sr-doped \cite{Nie2025PRL_crossover,Nie2025Sr} and pressurized thin films. \cite{HHWen2026pressure}

\textit{Fermi liquid.---}In the two-component picture, Fermi liquid behavior generally arises when the strongly correlated $d_{z^2}$ spins form a valence bond state that is decoupled from (or weakly coupled to) the weakly correlated $d_{x^2-y^2}$ electrons. Theoretically, this corresponds to a zero-energy gap in the $d_{x^2-y^2}$ self-energy due to the absence of electron scattering at low temperature. For fixed $U$ and other parameters, the gap size is proportional to $J$, corresponding to the energy required to break the VBS. For small $J$ or $U$, the interlayer valence bonds can be easily broken and the gap may even turn into a dip. This allows for hybridization between $d_{x^2-y^2}$ and $d_{z^2}$ quasiparticles even at half filling, leading to superconductivity with relatively low $T_c$ and a Fermi liquid normal state. This has indeed been observed in compressively strained thin films at ambient pressure. \cite{Hwang2025,Chen2025,Hwang2025FL}

\begin{figure}[t]
	\begin{centering}
		\includegraphics[width=0.6\textwidth]{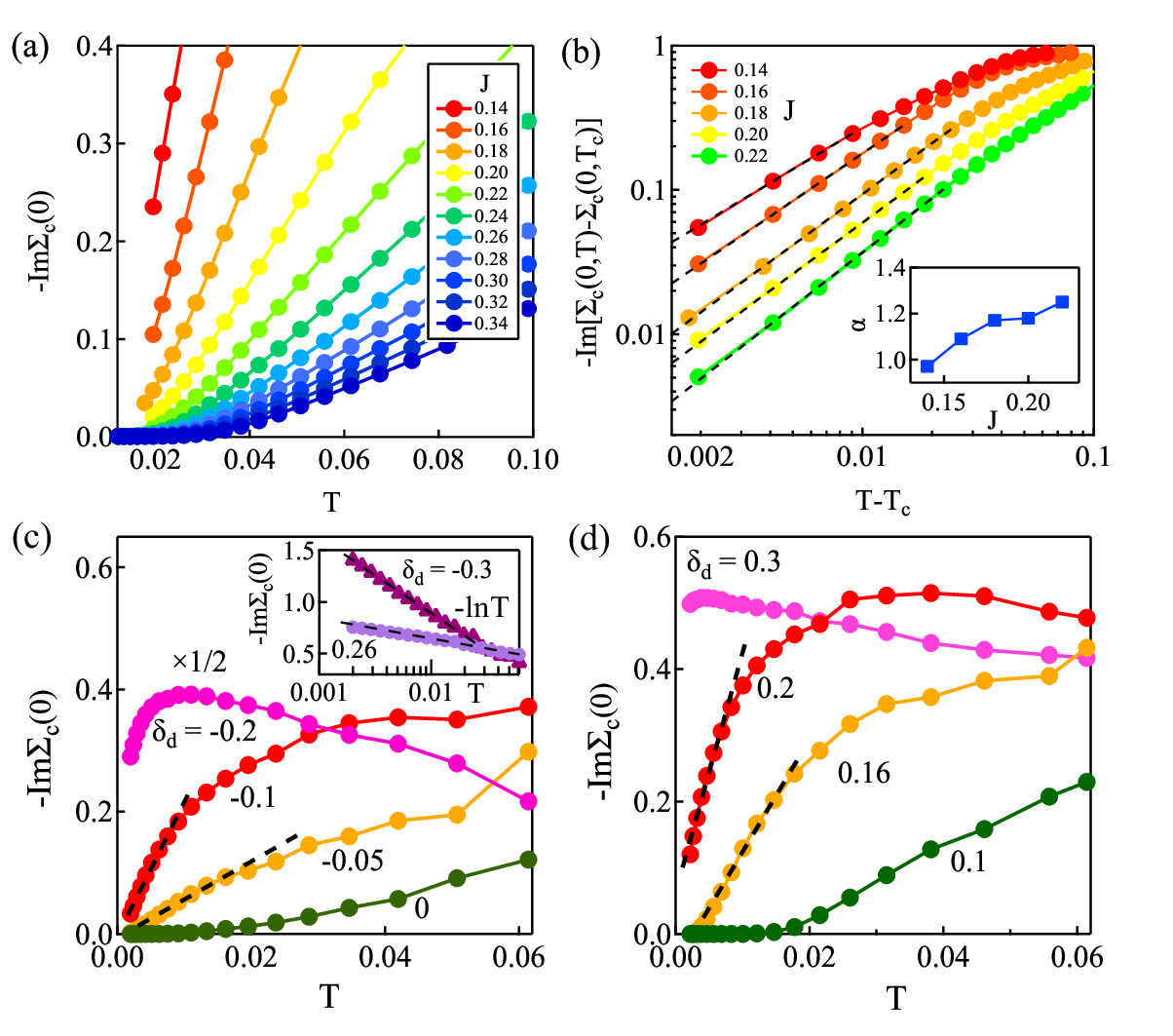}
		\par\end{centering}
	\caption{\textbf{Non-Fermi liquid.} (A) Temperature dependence of the imaginary part of the $d_{x^2-y^2}$ self-energy $-\text{Im}\Sigma_c(0)$ at $V=0.5$ for different $J$, obtained from Schwinger boson calculations of the $t$-$V$-$J$ model in the large-$U$ limit. \cite{Wang2025} (B) A log-log plot of the normalized self-energy $-\text{Im}\left[\Sigma_c(0,T)-\Sigma_c(0,T_c)\right]$ versus $T-T_c$ for different $J$. The dashed lines show power-law fittings of low-temperature data, $(T-T_c)^\alpha$. The inset shows $\alpha$ as a function of $J$. (C)(D) Temperature dependence of $-\text{Im}\Sigma_c(0)$ for different $d_{z^2}$ electron doping ($\delta_d<0$) or hole doping ($\delta_d>0$) at fixed $J=V=0.5$ and $U=7$. The inset of (C) shows the $-\ln T$ dependence at $\delta_d=-0.3$ and 0.26. Adapted from Refs. \cite{Wang2025,Wang2026}. }
	\label{fig:pseudNFL}
\end{figure}

\textit{Non-Fermi liquid.---}Strange metal behaviors in unconventional superconductors often indicate strong electronic correlations. In the Schwinger boson approach, it  naturally arises from electron-spinon-holon/doublon scattering (the hybridization vertex) when doping into the interlayer VBS, as illustrated in Figures \ref{fig1}B and \ref{fig2}A. By tuning the interlayer coupling $J$, Ref. \cite{Wang2025} obtained a quasi-linear-in-$T$ dependence of the imaginary part of the $d_{x^2-y^2}$ self-energy  ($-\text{Im}\Sigma_c(0)$) above optimal $T_c$, which gradually turns into a Fermi liquid as $J$ becomes too large to suppress the hybridization, as shown in Figures \ref{fig:pseudNFL}A-B. For fixed $J$ and finite $U$, Ref. \cite{Wang2026unified} found that the VBS gap of the Fermi liquid state around half filling is quickly suppressed by $d_{z^2}$ doping, and the enhanced hybridization vertex causes a NFL normal state roughly around the optimal $T_c$ for both hole and electron doping, as shown in Figure \ref{fig:pseudNFL}C-D, explaining the experimental observations in pressurized bulk \cite{HQYuan2024,Shen2025} and high-quality thin film bilayer nickelates. \cite{ZYChen60K} The NFL behavior was also obtained in DFT+DMFT calculations for the bulk, where the imaginary parts of both $d_{z^2}$ and $d_{x^2-y^2}$ self-energies exhibit linear-in-$T$ dependence at high pressure. \cite{YYCao2024}

\textit{Weakly insulating behavior.---}The weakly insulating (WI) region was observed above the superconducting phase in thin films when $T_c$ is gradually suppressed by Sr doping \cite{Nie2025PRL_crossover,Nie2025Sr} or increasing pressure. \cite{HHWen2026pressure} While Sr doping increases both the $d_{z^2}$ and $d_{x^2-y^2}$ hole doping, the hydrostatic pressure on thin films is estimated to shift the $\gamma$ pockets upward and increase the $d_{z^2}$ hole doping level. \cite{HHWen2026pressure} By increasing the $d_{z^2}$ doping ($|\delta_d|$), Ref. \cite{Wang2026unified} found that the NFL behavior gradually turns into weakly insulating where $-\text{Im}\Sigma_c(0)$ increases logarithmically with decreasing temperature, as shown in Figures \ref{fig:pseudNFL}C-D for $\delta_d= 0.26$ and -0.3. In this region, the $d_{x^2-y^2}$ self-energy exhibits a small peak at zero frequency, and the quasiparticle spectra disappear along the $k_x$ and $k_y$ axes, suggesting that the WI behavior is caused by incoherent Kondo scattering between $d_{x^2-y^2}$ electrons and $d_{z^2}$ moments when thermal fluctuations and doping destroy the VBS.  As $|\delta_d|$ further increases,  the $d_{z^2}$ electrons form well-defined quasiparticle bands that hybridize with $d_{x^2-y^2}$ electrons, leading to a hybridized Fermi liquid state accompanied by a Fermi surface reconstruction. The overall evolution is shown in Figure \ref{fig:doping} for $J=0.5$ and $J=0.1$ at $V=0.5$ and $U=7$. We see that the WI region on the hole doping side for $J=0.1$ requires much smaller $\delta_d$ compared to that of $J=0.5$, explaining why it has currently only been observed in thin films. 

\textit{Pseudogap.---}The two-component theory also provides a mechanism for the possible pseudogap phenomenon in bilayer nickelates, which has been suggested from several experimental observations. \cite{LYang2024,ZYChen2025,ZYChen2026Three} In cuprates, the pseudogap is often attributed to strongly fluctuating preformed Cooper pairs without achieving long-range phase coherence. In bilayer nickelates, the $d_{z^2}$ local spin singlet pairs as well as the hybridization-induced $d_{x^2-y^2}$ interlayer pairs play the role of preformed Cooper pairs, causing the pseudogap feature in the DOS within the temperature range $T_c < T< T_\Delta$, where $T_\Delta$ is the mean-field transition temperature of $d_{z^2}$ interlayer pairing. \cite{Wang2025}

\subsection{E. Pressure, oxygen content and Kondo effect}

Experimentally, bilayer nickelates are often studied under continuous tuning of external pressure or oxygen content. For theoretical comparison, it is important to investigate their effects and clarify how different model parameters may vary upon tuning these conditions. 

\textit{Pressure.---}The effect of hydrostatic pressure on the electronic structure has been studied in Refs. \cite{QHWang2025,DXYao2025DMRG,Verraes2025,GSu2025Unifying}. As expected, most hopping integrals increase monotonically with pressure, including the in-plane hopping of the $d_{x^2-y^2}$ orbital, the nearest-neighbor hybridization and the $d_{z^2}$ vertical hopping. Consequently, the overall bandwidth of Ni-$e_g$ orbitals increases under pressure, leading to a slightly weakened electronic correlation relative to the bandwidth. For bulk bilayer nickelates, hydrostatic pressure causes the structural transition and the emergence of the $\gamma$ Fermi pocket.  \cite{MWang2023Nature} Within the $I4/mmm$ phase, pressure further reduces the energy difference between the $d_{z^2}$ and $d_{x^2-y^2}$ orbitals, causing an upward shift of the $\gamma$ band and increased $d_{z^2}$ hole doping. \cite{DXYao2025DMRG,GSu2025Unifying,Osada2026,JingGuo2026}  The effect of pressure on compressively strained thin films has been studied in Refs. \cite{HHWen2026pressure,Osada2026}, revealing a similar upward shift of the $\gamma$ band and increased $d_{z^2}$ hole doping. For the interlayer magnetic coupling, naive estimates from $J/t\propto t_\perp^2/(Ut)$ suggest $J/t$ increases due to the increased ratio $t_\perp^2/t$ under pressure \cite{QHWang2025,DXYao2025DMRG}. By contrast, Ref. \cite{GSu2025Unifying} predicted a nonmonotonic evolution due to the modulation of apical oxygen $p_z$ energy level, and found it closely aligns with the $T_c$ curve of pressurized bulk La$_3$Ni$_2$O$_7$. RIXS measurements, on the other hand, reported nearly constant magnon bandwidth for ambient-pressure bulk and thin film bilayer nickelates under different compressive strains, implying that the effective magnetic energy scale may not change significantly under increasing compressive strain \cite{KJZhou2026}. Future investigations are needed to clarify this issue. The effect of chemical pressure has also been studied theoretically \cite{DXYao2025Nd,FYang2026Unified,CJWu2024CPL}, suggesting enhanced maximal $T_c$ by rare-earth elements substitution from La to Nd to Sm, possibly due to reduced interlayer distances (hence enhanced $J$). These results are qualitatively consistent with experimental observations \cite{JJZhang2025,MWang2026Sm,MWang2025,HLuo2026spin}. In the two-component theory, the rapidly enhanced $T_c$ and the NFL above optimal $T_c$ in pressurized bulk are mainly understood from the $d_{z^2}$ hole doping across the structural transition \cite{Wang2025}. However, the absence of weakly insulating behavior implies a limited range of $d_{z^2}$-$d_{x^2-y^2}$ charge transfer under pressure tuning, which may in turn explain the wide pressure range of the NFL state \cite{Wang2026unified}. The subsequent decrease of $T_c$ then reflects the gradual weakening of interlayer spin-singlet pairing under higher pressure. The enhanced maximal $T_c$ under chemical substitution and hydrostatic pressure follows the relation, $T_c^{\rm max}\approx 0.04-0.05 J$, predicted earlier for optimized conditions \cite{QQin2023}. By contrast, the ambient pressure superconductivity in thin films near $d_{z^2}$ half filling is attributed to their larger interlayer distance and hence smaller $J$ compared to the bulk \cite{Wang2026unified}. Further increasing $J$ may strengthen the $d_{z^2}$ valence bonds and suppresses hybridization-driven superconductivity, explaining the continuous decrease of $T_c$ upon Ln (lanthanide) substitution in La$_2$LnNi$_2$O$_7$/SLAO heterostructure. \cite{Osada2026}

\textit{Oxygen content.---}Many experiments have demonstrated the sensitivity of superconductivity to the oxygen content in bilayer nickelates. \cite{Hwang2026halfdome,ZYChen2025SIT,ZChen2024,YWang2025Interstitial,MWang2026Regulating,WLee2026Interlayer} It was shown that oxygen deficiency introduces strong scattering and inhomogeneity, which leads to a granular superconductor-to-insulator transition. \cite{Hwang2026halfdome,ZYChen2025SIT} The oxygen vacancies exhibit considerable spatial fluctuation but are observed to mainly occupy the inner apical O sites, \cite{ZChen2024} which reduce the $d_{z^2}$ interlayer superexchange and suppress the interlayer pairing superconductivity. Theoretically, this has been supported by first-principles calculations in Ref. \cite{BHuang2024}, which found that the removal of the inner apical oxygen is energetically favorable. In addition, Ref. \cite{FYang2023} studied the effect of randomly distributed oxygen vacancies using the real-space RPA, and found that the superconductivity is suppressed by oxygen vacancies. Moreover, Ref. \cite{FYang2023} suggested the emergence of local moments in the vicinity of apical oxygen vacancies that are also harmful to the superconductivity. 

\begin{figure}[t]
	\begin{centering}
		\includegraphics[width=0.6\textwidth]{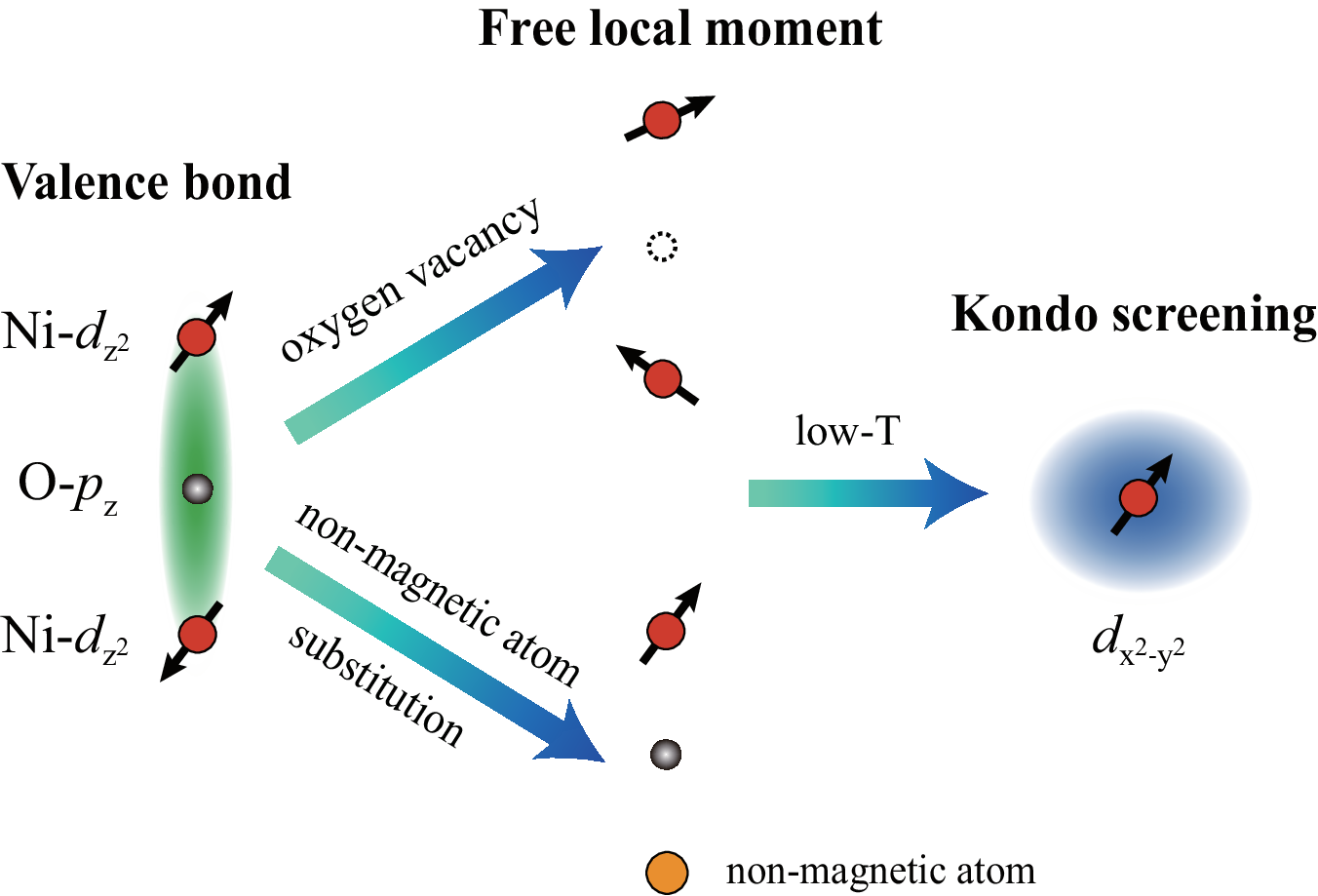}
		\par\end{centering}
	\caption{\textbf{Kondo effect.} Schematic illustration of Kondo effect induced by inner apical oxygen vacancies or substitution of Ni by non-magnetic ions. }
	\label{fig:Kondo}
\end{figure}

Another consequence of oxygen vacancies related to the local moment formation is the {\it Kondo effect} observed in some non-superconducting samples. \cite{HHWen2024Kondo,KuiJin2026,KeWang2026,Kumar2026} In the $t$-$V$-$J$ model, an apical oxygen vacancy breaks the local vertical $d_{z^2}$ valence bond and leads to two decoupled $d_{z^2}$ local moments. The same effect may also arise from substitution of Ni with non-magnetic atoms such as Al,\cite{GHCao2026} as shown schematically in Figure \ref{fig:Kondo}. These $d_{z^2}$ moments are surrounded by itinerant $d_{x^2-y^2}$ electrons, and the hybridization between them causes an effective Kondo scattering term. \cite{Wang2026unified} As temperature decreases, the Kondo screening effect takes place and leads to $-\ln T$ resistivity and negative magnetoresistivity, as observed experimentally in non-superconducting samples  \cite{HHWen2024Kondo,KuiJin2026,KeWang2026,Kumar2026} and explained in a recent study of the two-component theory. \cite{Wang2026unified}

When oxygen vacancies are eliminated by increasing the oxygen content, both the interlayer coupling and global phase coherence are enhanced, leading to bulk  superconductivity with increased $T_c$. Moreover, increasing oxygen content also introduces hole doping to the Ni-3$d$ orbitals, which promotes Ni-3$d_{z^2}$ metallization and increases the transition temperature of hybridization-driven superconductivity. This was indeed supported by recent XAS and RIXS measurements, which found that oxygenation enhances the $d_{z^2}$ orbital delocalization and drives superconductivity. \cite{KJZhou2026}  On the other hand, overdoped oxygen atoms are harmful to superconductivity for two reasons: i) They induce too many $d_{z^2}$ holes that strongly suppress the interlayer magnetic correlation; \cite{Wang2026unified} ii) They may reside on interstitial sites and form stripe orders that suppress $T_c$. \cite{Hwang2026halfdome,YWang2025Interstitial} Overall, tuning oxygen content leads to a dome-like structure in bilayer nickelates as reported in Refs. \cite{Hwang2026halfdome,Nie2025PRL_crossover,Nie2025Sr}.

\subsection{F. Other pairing theories} 

In this subsection, we briefly review other  pairing theories of bilayer nickelates that have not been covered in detail, including weak correlation theories, numerical simulations and others. 

\textit{Weak correlation theories.---}In contrast to the strong correlation theories discussed in previous sections, the weak correlation theories assume an itinerant picture for both Ni-$e_g$ orbitals, and take tight-binding Hamiltonians from first-principles calculations as their starting points. The multiorbital electron interactions are then treated in a perturbative way via approaches like the random-phase approximation (RPA) \cite{DXYao2023,FYang2023,Dagotto2024,Dagotto2025RPA,HHChen2025,Savrasov2024,Eremin2023,Braz2025PRR,FYang2025Band,WQChen2025,Raghu2026,HHChen2026Nearly} and functional renormalization group (FRG). \cite{QHWang2023,QHWang2025,QHWang2026Strain,QHWang2026Tunable,JPHu2025,JPHu2025PRL,JPHu2025Opposite} The resulting spin fluctuations associated with certain Fermi surface nestings then serve as the pairing glue and cause superconductivity. Most of these theories predict $s^{\pm}$-wave pairing gaps on $\alpha$, $\beta$ and $\gamma$ bands, \cite{FYang2023,Dagotto2024,Dagotto2025RPA,WQChen2025,QHWang2023,QHWang2025,QHWang2026Strain,QHWang2026Tunable,JPHu2025PRL,JPHu2025Opposite} and take the view that the $\gamma$ hole pocket is important for the $s^{\pm}$-wave superconductivity since the strongest spin fluctuation is associated with the nesting between $\gamma$ and $\beta$ pockets. \cite{FYang2023,QHWang2023,WQChen2025} A different $s^{\pm}$-wave pairing without the $\gamma$ hole pockets was proposed in Ref. \cite{FYang2025Pairing}, in which case the pairing is driven by nesting between the $\alpha$ and $\beta$ pockets. In a recent RPA study, the Hund's coupling was also shown to control the competition between $s$-wave and $d$-wave pairings when the $\gamma$ pocket is absent.\cite{Raghu2026} For bulk hole-doped bilayer nickelate La$_{3-x}$Sr$_x$Ni$_2$O$_7$, Ref. \cite{HHChen2026Nearly} recently predicted ambient-pressure $d_{x^2-y^2}$-wave superconductivity driven by enhanced spin fluctuations associated with nearly perfect nesting of the $\gamma$ pocket at $x\approx 0.4$. Therefore, different from the robust $s^{\pm}$-wave pairing in strong correlation theories, weak correlation theories also predict competing $d_{xy}$ \cite{Eremin2023,HHChen2025,Savrasov2024,Braz2025PRR,FYang2025Band} or $d_{x^2-y^2}$ \cite{Eremin2023,JXLi2025,HHChen2026Nearly} pairing symmetries, whose stabilization depends sensitively on the details of crystal field splitting, \cite{HHChen2025} interlayer Coulomb interaction, \cite{Braz2025PRR,JXLi2025} or doping level. \cite{FYang2025Band,HHChen2026Nearly} It should be mentioned that Ref. \cite{JPHu2025PRL} investigated the effect of electron-phonon coupling by combining first-principles calculation with FRG, and found the high $T_c$ of  $s^{\pm}$-wave pairing superconductivity is driven by cooperation between electronic correlation and out-of-plane electron-phonon coupling. However, whether the electron-phonon coupling is truly relevant for the nickelate superconductivity remains debated. \cite{ZYLu2024EPC,QHWang2025,Talantsev2024,Li2025SciBull}

\textit{Numerical methods and others.---}Ref. \cite{QHWang2025VMC} studied the bilayer two-orbital Hubbard model and extended $t$-$J$ model via the variational quantum Monte Carlo method, and found an $s^{\pm}$-wave pairing dominated by the $d_{z^2}$ orbital. Similar $d_{z^2}$ dominated $s^{\pm}$-wave pairing has been obtained by many other theoretical studies, including dynamic cluster approximation quantum Monte Carlo simulations \cite{Dagotto2026Interlayer} and DMRG calculations \cite{Kuroki2024Pair} of the bilayer two-orbital Hubbard model, renormalized mean-field theory approach to the bilayer two-orbital $t$-$J$ model, \cite{DXYao2023tJ,DXYao2025pairing} and other self-consistent methods. \cite{Kuroki2024,TZhou2023Impurity} Quantum Monte Carlo simulations of bilayer Hubbard models show a doping-driven Mott-insulator-superconductor transition, \cite{Ma2022Doping} and identify Hund's coupling and crystal-field splitting as the key factors controlling the competition of $d$-wave and $s^{\pm}$-wave pairings. \cite{Xiong2026} Tensor-network methods like DMRG \cite{Kuroki2024DMRG,GSu2025,GMZhang2023,Bohrt2024Feshbach,FYang2026NC} and iPEPS \cite{WLi2024,GSu2024} have been used to study different types of $t$-$J$ models for bilayer nickelates.  For example, Ref. \cite{FYang2026NC} studied a $t$-$J$ model using both mean-field theory and DMRG calculations, and proposed that a perpendicular electric field may suppress the interlayer $s$-wave pairing and lead to an intralayer $d$-wave pairing with higher $T_c$. Neural quantum state calculations of the mixed-dimensional bilayer $t$-$J_\parallel$-$J_\perp$ model have also found robust superconductivity over a wide doping and coupling ranges.  \cite{Lange2026} Besides the bilayer two-orbital models, there are also theoretical studies explicitly considering the oxygen $p$ orbitals. \cite{WWu2024SCPMA,DXYao2025DQMC} 

Due to the limited length and the major focus of this paper, we have not covered in detail the theoretical investigations focusing on the density waves,  \cite{Werner2023,JHu2024electronic,Raghu2025Origin,Zhang2025Spin,Botana2025Assessing,KCao2025,MJiang2025Assessing,Sawatzky2025,Leonov2025,Lu2025Charge,KJiang2026Itinerant,QSi2026Magnetic}  cuprate-like intralayer $d$-wave pairing mechanisms \cite{KJiang2024,TXiang2024,WKu2024,FCZhang2025} and mechanisms of trilayer nickelates. \cite{Leonov2024Trilayer,QHWang2024Trilayer,Li2024Effective,QQin2024}

\section{IV. SUMMARY AND OUTLOOK}

We have provided a brief survey of recent progress on the superconducting mechanism of bilayer nickelate superconductors, with a focus on strong correlation interlayer pairing theories. Starting from key experimental observations, we distilled the essential physical ingredients and effective theoretical models. The minimal bilayer  two-orbital $t$-$J$ model including the interorbital hybridization and Hund's coupling  serves as the fundamental framework for understanding the low-energy physics. Starting from the atomic-limit interlayer valence bond picture of the nearly half-filled $d_{z^2}$ orbitals, we introduce different pairing mechanisms by considering different limits. In particular, we have focused on the two-component scenario for mobilizing the local interlayer pairs into a coherent superconducting state through interorbital hybridization.

\begin{figure}[t]
	\begin{centering}
		\includegraphics[width=0.7\textwidth]{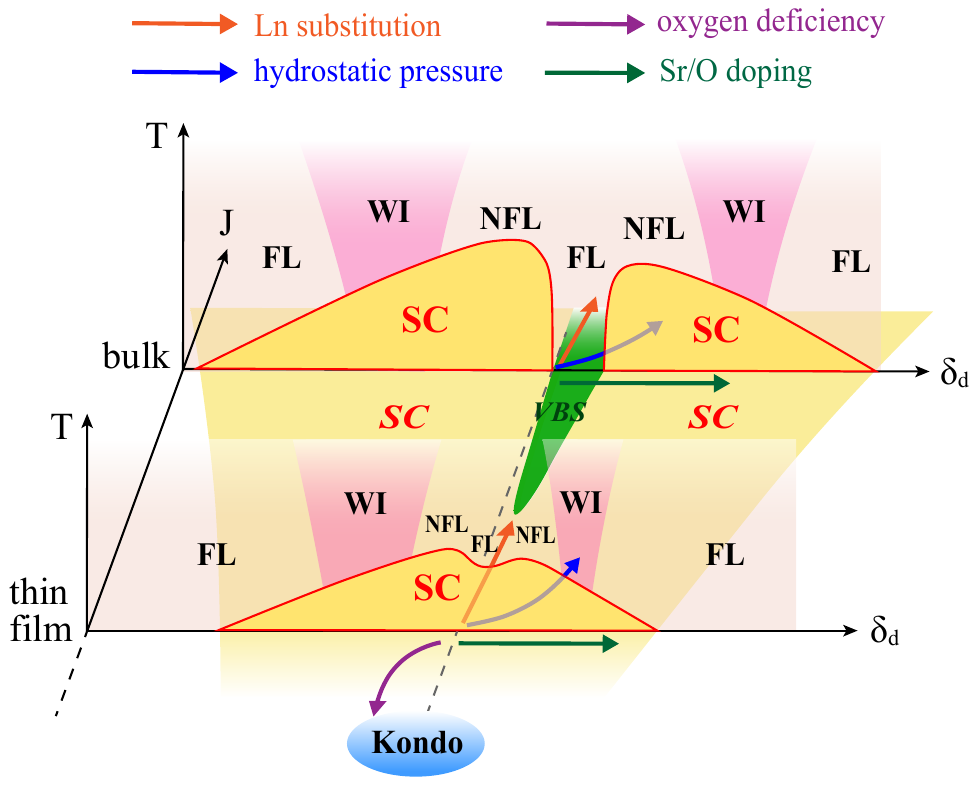}
		\par\end{centering}
	\caption{\textbf{Schematic phase diagram.} The two-component theory predicts two key parameters, the interlayer superexchange $J$ and the $d_{z^2}$ hole doping $\delta_d$, controlling the superconductivity and normal state properties. The large and small $J$ correspond to the bulk and compressively strained thin film bilayer nickelates, respectively. FL, NFL and WI mark different normal states above the SC or VBS ground states. Experimental studies listed in Table \ref{tab:experiments} are represented by different routes (arrows) in this phase diagram. Adapted from Ref. \cite{Wang2026unified}. }
	\label{fig:phase}
\end{figure}

\begin{table*}[tbp]
	\centering
	\footnotesize
	\caption{Representative experiments on bilayer nickelate superconductors mapped onto the $(\delta_d,\, J)$ phase diagram of Fig.~\ref{fig:phase}. $\delta_d$ denotes the Ni-$3d_{z^2}$ hole doping and $J$ is the interlayer superexchange interaction. FL, NFL, and WI represent the Fermi-liquid, non-Fermi-liquid and weakly insulating normal states. HP, AP, and h.p. denote high, ambient, and hydrostatic pressures, respectively. MR stands for magnetoresistance.}
	\label{tab:experiments}
	\setlength{\tabcolsep}{4pt}
	\begin{tabular}{@{}>{\raggedright\arraybackslash}p{3.1cm}>{\raggedright\arraybackslash}p{2.5cm}>{\raggedright\arraybackslash}p{3.5cm}>{\raggedright\arraybackslash}p{2.7cm}>{\raggedright\arraybackslash}p{3.0cm}@{}}
		\toprule
		Material & Tuning parameter & Position on $(\delta_d,J)$ diagram & Maximum $T_c$ \newline (K) & Normal state \\
		\midrule
		La$_3$Ni$_2$O$_7$ bulk single crystal & hydrostatic pressure (h.p.)  & large $J$; \newline $\delta_d>0$ at HP \cite{MWang2023Nature} & $\sim80$\cite{MWang2023Nature} & NFL ($\Delta \rho\propto T$)\cite{HQYuan2024} \\
		\midrule
		(La,Sm)$_3$Ni$_2$O$_7$ bulk single crystal & chemical pressure + h.p. & shortens $c$ $\to$ larger $J$;\newline $\delta_d>0$ at HP \cite{DfDuan2026}  & 96\cite{JJZhang2025}  & NFL ($\Delta \rho\propto T$)\cite{JJZhang2025} \\
		\midrule
		(La,Nd)$_x$Ni$_2$O$_7$ \newline bulk polycrystal & chemical pressure + h.p. & shortens $c$ $\to$ larger $J$;\newline $\delta_d>0$ at HP \cite{DXYao2025Nd} &  $\sim 100$\cite{MWang2025}  & \\
		\midrule
		La$_2$PrNi$_2$O$_7$/SLAO \newline thin film & compressive strain + ozone oxidation & small $J$;\newline $\delta_d\approx 0$ (no $\gamma$ pocket)\cite{ZXShen2025} & 42 -- 48 \cite{Hwang2025,Hwang2025b} & FL ($\frac{1}{\rho}=\frac{1}{\rho_0+AT^2}+c$)\cite{Hwang2025b} \\
		\midrule
		(La,Pr)$_3$Ni$_2$O$_7$/SLAO \newline thin film & compressive strain + ozone oxidation & small $J$;\newline $\delta_d>0$ ($\gamma$ pocket)\cite{QKXue2025}  & 45\cite{Chen2025}  & FL ($\Delta \rho\propto T^2$)\cite{Chen2025}  \\
		\midrule
		(La,Pr)$_3$Ni$_2$O$_7$/SLAO  \newline high-quality thin film  & compressive strain + ozone oxidation & small $J$;\newline $\delta_d>0$ ($\gamma$ pocket)\cite{ZYChen2026} & 38 -- 63\cite{ZYChen60K} & $\Delta \rho\propto T^{\alpha}$:\newline $\alpha\approx 1.1$ (high $T_c$) $\alpha\approx 1.8$ (low $T_c$)\cite{ZYChen60K} \\
		\midrule
		La$_{3-x}$Sr$_x$Ni$_2$O$_{7-\delta}$/ \newline SLAO thin film & Sr doping ($x\le0.21$) + oxygen content & small $J$;\newline scans $\delta_d$ across dome & $\sim 40$\cite{Nie2025PRL_crossover}  & metallic $\to$ WI ($\rho\sim -\ln T$) as hole doping increases \cite{Nie2025PRL_crossover} \\
		\midrule
		La$_{3-x}$Sr$_x$Ni$_2$O$_{7-\delta}$/ \newline SLAO thin film & Sr doping ($x\le0.45$) & small $J$;\newline Sr doping increases $\delta_d$ & $\sim42$ ($x=0.09$) \cite{Nie2025Sr} & FL $\to$ WI ($x\ge0.37$) as $x$ increases \cite{Nie2025Sr} \\
		\midrule
		La$_{3-x}$Pr$_x$Ni$_2$O$_7$ \newline /SLAO thin film & compressive strain + h.p. & h.p. increases $J$ and $\delta_d$ ($\gamma$ pocket expands)\cite{HHWen2026pressure} & 30 (AP) \newline 48.5 (7 GPa)\newline 42.8 (13 GPa)\cite{HHWen2026pressure} & FL ($\alpha\approx 2$) $\to$ NFL ($\alpha<2$) $\to$ WI ($-\ln T$) as pressure increases\cite{HHWen2026pressure} \\
		\midrule
		La$_2$LnNi$_2$O$_7$/SLAO \newline thin film (Ln = lanthanide) & Ln substitution at AP;\newline h.p. for Ln = Sm,Nd samples & Ln substitution shortens $c$ (increases $J$);\newline h.p. increases $J$ and $\delta_d$\cite{Osada2026} & $T_c$ decreases as $c$ shortens (AP); \newline h.p. increases $T_c$: 41--42 (AP) \newline 67--73 (16 GPa)\cite{Osada2026} & $\Delta \rho\propto T^{\alpha}$:\newline $\alpha\approx 1.2$ (high $T_c$) $\alpha\approx 1.8$ (low $T_c$)\cite{Osada2026} \\
		\midrule
		La$_3$Ni$_2$O$_{7-\delta}$/LAO thin film &   hole doping + oxygen content (ionic-liquid gating) & apical oxygen vacancies break valence bonds $\to$ free $d_{z^2}$ moments; \newline hole doping increases $\delta_d$  & non-SC & Kondo ($\Delta\rho\propto -\ln T$, negative MR) at low doping \cite{KuiJin2026} \\
		\midrule
		La$_3$Ni$_2$O$_{7-\delta}$/LAO thin film & oxygen content & apical oxygen vacancies break valence bonds $\to$ free $d_{z^2}$ moments & non-SC & Kondo ($\Delta\rho\propto -\ln T$, negative MR) \cite{HHWen2024Kondo} \\
		\midrule
		(La,Pr)$_3$Ni$_2$O$_{7-\delta}$ thin films on LAO and SLAO substrates & compressive strain + oxygen content & strain/oxygenation delocalizes $d_{z^2}$ orbital (increases $\delta_d$)\cite{KJZhou2026} & $\sim50$ (SLAO substrate) & $d_{z^2}$ metallization suppresses SDW \cite{KJZhou2026} \\
		\midrule
		\bottomrule
	\end{tabular}
\end{table*}

\subsection{A. Experimental comparisons and schematic phase diagram}

The two-component scenario yields numerous theoretical predictions that are consistent with current experimental observations on both bulk and thin-film bilayer nickelates. These are summarized as follows:

1) Anisotropic $s^{\pm}$-wave pairing with gap minima along the zone diagonal on the $\alpha$ and $\beta$ pockets and an isotropic gap on the $\gamma$ pocket, \cite{YFYang2023,QQin2023,Wang2025} consistent with current experimental observations. \cite{HHWen2025,ZYChen2025,ZYChen2026Three} The predicted gap ratio,  $2\Delta^\text{max}/T_c\approx 7.5-9$, \cite{QQin2023} has been confirmed by recent ARPES measurements. \cite{ZYChen2026Three}

2) Crucial role of $d_{z^2}$-orbital metallization for the superconductivity. For large $J$ as in bulk, this predicts enhanced $T_c$ by small $d_{z^2}$ hole doping away from half filling and a highly asymmetric superconducting dome \cite{Wang2025,Wang2026unified}. For small $J$ as in thin films, the $d_{z^2}$ orbitals can be metallized even near half filling, explaining why thin films can be superconducting both with \cite{QKXue2025} and without \cite{ZXShen2025} the $\gamma$ pocket. For heavier $d_{z^2}$ doping, the pairing energy scale is suppressed, leading to decreased $T_c$, consistent with experiments of thin films under Sr doping \cite{Nie2025PRL_crossover,Nie2025Sr}, oxygenation \cite{Hwang2026halfdome,Nie2025PRL_crossover} or pressure tuning \cite{HHWen2026pressure}. 

3) A maximal ratio $T_c/J\sim 0.04-0.05$ under optimal conditions, \cite{QQin2023,QQin2025} which gives the correct magnitudes of $T_c$ for bulk and thin film bilayer nickelates when using experimentally estimated $J$. For trilayer nickelates, it predicts a reduced maximal ratio $T_c/J\sim 0.02-0.03$, \cite{QQin2024} which is opposite to that of cuprates but consistent with nickelate experiments. The increase of maximal $T_c$ with $J$ has been confirmed experimentally by chemical substitution that reduces the interlayer distance in pressurized bulk. \cite{JJZhang2025,MWang2025,MWang2026Sm}   For thin films near $d_{z^2}$ half filling, it predicts a maximal $T_c$ of about 50 K at ambient pressure and a nonmonotonic $T_c-J$ relation, consistent with experimental observation with rare-earth substitution \cite{Osada2026}.

4) Non-Fermi liquid behavior induced by strong interorbital scattering, \cite{Wang2025,Wang2026unified,YYCao2024} which turns into a FL weakly coupled to the VBS around half-filling, or into a weakly insulating state for relatively large $d_{z^2}$ hole doping, consistent with experiments on Sr-doped \cite{Nie2025PRL_crossover,Nie2025Sr} and pressurized thin films. \cite{HHWen2026pressure} For large interlayer coupling $J$, the WI region requires heavier $d_{z^2}$ doping, which explains its absence in bulk experiments so far. The two-component theory also predicts a pseudogap phenomenon associated with preformed Cooper pairs, \cite{Wang2025} whose signature has been detected in experiments. \cite{LYang2024,ZYChen2025,ZYChen2026Three}

5) Kondo scattering by $d_{z^2}$ local moments, along with the simultaneous suppression of interlayer Cooper pairing caused by inner apical oxygen vacancies or non-magnetic substitution of Ni ions, \cite{Wang2026unified} which explains the observations in non-superconducting samples of bilayer nickelates. \cite{HHWen2024Kondo,KuiJin2026,KeWang2026,Kumar2026}

Table \ref{tab:experiments} summarizes some of the above representative experiments. Unlike other theories, the two-component theory identifies the interlayer superexchange $J$ and the $d_{z^2}$ hole doping $\delta_d$ as the key parameters controlling superconductivity in bilayer nickelates. This leads to the schematic phase diagram shown in Fig.~\ref{fig:phase}. Large and small $J$ correspond to the bulk and compressively strained thin films, respectively. The FL, NFL, and WI normal states are also indicated above the superconducting (SC) or VBS ground states. The arrows illustrate how these two parameters may be tuned under different experimental conditions. Hydrostatic pressure may simultaneously induce $d_{z^2}$ hole doping $\delta_d$ and enhance the interlayer superexchange $J$; rare-earth (Ln) substitution mainly tunes the interlayer distance and hence $J$;  Sr doping or excess O increases the $d_{z^2}$ hole concentration $\delta_d$; and oxygen deficiency breaks the interlayer pairs and introduces electron doping, giving rise to free $d_{z^2}$ local moments and incoherent Kondo scattering. The phase diagram thus provides a unified framework for understanding the major experimental observations of superconductivity and normal-state properties in bulk and thin film bilayer nickelates.

\subsection{B. Open questions}

Despite of the above successes, important open questions remain to be addressed as listed below:

1) Quantitative comparisons with experiments, including the detailed pressure and doping dependences of $T_c$, are still lacking. Refined calculations on more sophisticated models (incorporating first-principles parameters, in-plane correlations, Hund's rule coupling, etc.) are needed. 

2) Investigations of the two-component theory using other analytical and numerical methods are demanding. It is always challenging to solve a strongly correlated model. Approximations cannot be avoided, but the conclusions require cross-check by different approaches. 

3) The nature of the spin/charge density waves and their microscopic connection to superconductivity remain to be clarified. Within the two-component framework, one may attribute the density waves to the interplay among interlayer superexchange, the Ruderman-Kittel-Kasuya-Yosida (RKKY) interaction (mediated by $d_{z^2}$-$d_{x^2-y^2}$ hybridization), and the Hund-induced double-exchange interaction. \cite{QSi2026Magnetic} More dedicated studies are needed to explore this direction. 

4) Some of our theoretical results call for additional experimental confirmation. For instance, a decisive experimental determination of the pairing symmetry and a systematic experimental investigation of how $T_c$ varies with $J$ and $d_{z^2}$ occupation have not been available. Likewise, the nature of the weakly insulating normal state at some $d_{z^2}$ hole dopings requires further theoretical and experimental elucidation. 

5) In contrast to the two-component theory, weak correlation theories often rely heavily on Fermi surface topology and hence on the $d_{x^2-y^2}$ occupation, whereas Hund-driven pairing can be suppressed by $d_{z^2}$ hole doping and thus may be strongly influenced by the $\gamma$ pocket. Careful experimental scrutiny of these distinct predictions may help identify the primary pairing mechanism and establish the key conditions for maximizing $T_c$ in thin films or for achieving ambient-pressure superconductivity in bulk crystals. 

6) The multi-orbital model developed here for bilayer nickelates may be modified and extended to other correlated systems, such as the infinite-layer nickelates, the trilayer nickelates, and the $f$-electron superconductors with valence fluctuations. In-depth investigations of the differences and similarities of these systems may lead to a unified framework for understanding multi-orbital superconductors.  

Addressing these open challenges and unsolved questions is essential for a comprehensive understanding of the bilayer nickelate physics, which may eventually lead to new guiding principles for material design of more high-$T_c$ superconductors.

\section*{Acknowledgments}
The authors thank Q. Qin, Y. Cao, G.-M. Zhang, F.-C. Zhang, and many other colleagues for stimulating discussions. This work was supported by the National Natural Science Foundation of China (Grants No. 12304174 and No. 12474136). The funders had no role in study design, data collection and analysis, decision to publish, or preparation of the manuscript.

\end{document}